\documentclass[a4paper,12pt]{article}
\usepackage{amssymb, amsmath}
\usepackage[english]{babel}
\usepackage[applemac]{inputenc}
\usepackage[dvips]{graphicx}
\usepackage{amsfonts}
\usepackage{mathrsfs}
\usepackage{color}
\newcommand{\rd}{\textcolor{red}}

\newcommand{\ud}{\mathrm{d}}
\newcommand{\shalf}{{\hbox{\small{$\frac{1}{2}\hskip 0.01cm$}}}}
\newtheorem{theorem}{Theorem}[section]

\newtheorem{lemma}{Lemma}[section]
\newtheorem{remark}{Remark}[section]
\newtheorem{corollary}{Corollary}[section]

\begin{document}
{\qquad \hfill \rd{\textbf{DRAFT:  JStatPhys, \textit{{May 28, 2009}}}}}

\begin{center}

{\Large{\textbf{On the nature of Bose-Einstein condensation \\
in disordered systems}}}

\vspace{1cm}

\textbf{Thomas Jaeck\footnote{PhD student at UCD and Universit\'{e} de la M\'{e}diterran\'{e}e
(Aix-Marseille II, France), \\
e-mail: Thomas.Jaeck@ucdconnect.ie, phone: +353 1 7162571 },
Joseph V. Pul\'{e}\footnote{e-mail: joe.pule@ucd.ie, phone: +353 1 7162568}}\\
School of Mathematical Sciences, University College Dublin\\
Belfield, Dublin 4, Ireland

\vspace{1cm}

\textbf{Valentin A. Zagrebnov\footnote{e-mail: Valentin.Zagrebnov@cpt.univ-mrs.fr, phone: +33 491 26 95 04}}\\
Universit\'{e} de la M\'{e}diterran\'{e}e (Aix-Marseille II), \\ Centre de Physique Th\'{e}orique - UMR 6207
CNRS,  Luminy - Case 907 \\ 13288 Marseille, Cedex 09, France

\vspace{1cm}

\begin{abstract}
\noindent {We study the perfect Bose gas in \textit{random} external potentials and show that there is
generalized
Bose-Einstein condensation in the random eigenstates if and only if
the same occurs in the one-particle kinetic-energy eigenstates, which corresponds to the generalized
condensation
of the free Bose gas. Moreover, we prove that the amounts of both condensate densities are equal.
Our method is based on the derivation of an explicit formula for the occupation measure in the one-body
kinetic-energy eigenstates which describes the repartition of particles among these non-random states.
This technique can be adapted to re-examine
the properties of the perfect Bose gas in the presence of weak (scaled) \textit{non-random} potentials,
for which we
establish similar results.}
\end{abstract}

\end{center}


\vspace{0.5cm}

\textbf{AMS:} 82B10, 82B23, 81V70

\newpage
\tableofcontents
\section{Introduction}
\setcounter{equation}{0}
\renewcommand{\theequation}{\arabic{section}.\arabic{equation}}

The study of Bose-Einstein Condensation (BEC) in random media has been an important area for a long time,
starting with the papers by Kac and Luttinger, see \cite{KL1}, \cite{KL2},  and then by Luttinger and
Sy \cite {LS}. In the last reference, the authors studied a non-interacting (\textit{perfect}) one
dimensional system
with point impurities distributed according to the Poisson law, the so-called \textit{Luttinger-Sy model}.
The authors
conjectured a macroscopic occupation of the random ground state, but this was not rigorously proved until
\cite{LZ}. {Although}
the \emph{free} Bose gas (i.e., the \textit{perfect} gas without external potential) does not exhibit
BEC for dimension
less than three, the randomness can enhance BEC even in \textit{one dimension}, see e.g. \cite{LPZ}.
This striking
phenomenon is a consequence of the exponential decay of the one particle density of states at the
bottom of the spectrum,
known as \emph{Lifshitz tail }, or \lq\lq doublelogarithmic" asymptotics,
which is generally believed to be associated with the existence of localized eigenstates \cite{PF}.

BEC, however, is usually associated with a macroscopic occupation of the lowest one-particle
kinetic-energy eigenstates,
which are spatially extended (plane waves). Therefore, it is not immediately clear whether the
phenomenon discovered
in random boson gases, i.e. macroscopic occupations of localized one-particle states, has any
relation to the standard
BEC. This is of particular interest in view of the well-known Bogoliubov approximation \cite{ZB} and
its applications to
disordered boson systems, see e.g. \cite{HM}, \cite{KT}, where the \emph{a priori} assumption of the
momentum-space
condensation is essential, and is far from {trivial} to check.

In this paper, we prove that for the perfect Bose gas {in} a general class of non-negative
random potentials,
BEC in the random localized one-particle states and BEC in the lowest one-particle kinetic-energy
states occur
simultaneously, and moreover the density of the condensate fractions are equal. Our line of reasoning
is also applicable
to some non-random systems, for example to the case of the perfect gas in weak (scaled) external
potentials studied
in \cite{P}.

The structure of the paper is as follows: in Section \ref{model-notations} we describe our {disordered
system}, and in
Section \ref{section-gBEC-random-eigenstates}, we recall standard results about the corresponding
\emph{perfect} Bose gas.
The existence of \textit{generalized} BEC in the eigenstates of the one-particle Schr\"{o}dinger operator
follows from
the finite value of the critical density for \emph{any dimension}, which is a consequence of the Lifshitz
tail in the
limiting Integrated Density of States (IDS). It is well-known that the IDS is a \textit{non-random}
quantity, see e.g.
\cite {PF}, and therefore the BEC density is also non-random in the thermodynamic limit. In Section
\ref{Sect-kinetic-occupation-measure}, we turn to the main result of this paper: we show that this
phenomenon occurs
\textit{if and only if} there is also occupation of the lowest one-particle kinetic-energy eigenstates.
The latter
corresponds to the usual generalized BEC in the \emph{free} Bose gas, that is a perfect gas without
external potential.
To establish this we prove the existence of a non-random limiting occupation measure for kinetic energy
eigenstates, and
moreover, we obtain an explicit expression for it. {To this end,} we need some estimates for the IDS
before the
{thermodynamic} limit, namely a \emph{finite volume} version of the Lifshitz tail estimates, which we
prove in Section
\ref{Sect-Lifshitz-tails}, using techniques developed in \cite{KM}, \cite{S}. For any \emph{finite} but
large enough
system, these bounds hold almost surely with respect to random potential {realizations}. In Section
\ref{Sect-g-BEC-Lutt-Sy}, we look at the particular case of the Luttinger-Sy model and examine the nature
of the
condensate in the one-particle kinetic energy eigenstates, showing that although there is generalized BEC,
\textit{no condensation} occurs in any of them. In Section \ref{Sect-appl-weak-potentials},
we describe briefly how the method developed in Section \ref{Sect-kinetic-occupation-measure} applies with
minor
modifications to a perfect Bose gas in a general class of weak (scaled), non-random external potentials.
To make the
paper more accessible and easy to read, we postpone some technical estimates concerning
random potentials and
Brownian motion to Appendices A and B, respectively.
\section{Model, notations and definitions}\label{model-notations}

Let {$\{\Lambda_{l}:= (-l/2, l/2)^{d}\}_{l \geqslant 1}$ be a sequence of hypercubes of side $l$ in
$\mathbb{R}^{d}, d \geqslant 1$, centered at the origin of coordinates with volumes $V_{l} = l^{d}$.}
We consider a system of identical bosons, of mass $m$, contained
in $\Lambda_{l}$. For {simplicity}, we use a system of units such that $\hbar = m = 1$. First we define the
self-adjoint one-particle kinetic-energy operator of our system by:
\begin{eqnarray}\label{Kinetic-energy-operator}
h_{l}^{0} := -\shalf \Delta_D,
\end{eqnarray}
acting {in} the Hilbert space $\mathscr{H}_{l} := L^{2} (\Lambda_{l})$. The subscript $D$ stands for
\textit{Dirichlet}
boundary conditions. We denote by $\{\psi_{k}^{l}, \varepsilon_{k}^{\l}\}_{k \geqslant 1}$ the set of
normalized
eigenfunctions and eigenvalues corresponding to $h_{l}^{0}$. By convention, we order the eigenvalues
(counting the multiplicity) as $\varepsilon_{1}^{l} \leqslant  \varepsilon_{2}^{l} \leqslant
\varepsilon_{3}^{l}\dots\,\,$.

We define an external random potential $v^{(\cdot)}(\cdot)\, : \,\Omega \times \mathbb{R}^{d}
\rightarrow \mathbb{R}, \ x \mapsto v^{\omega}(x)$ as a random field on a probability space
$(\Omega, \mathcal{F}, \mathbb{P})$, satisfying the following conditions:\\
(i) $v^{\omega}, \omega\in \Omega$, is non-negative;\\
(ii) $p := \mathbb{P} \{\omega: v^{\omega}(0) = 0\} < 1$.\\
As usual, we assume that this field  is \textit{regular}, \textit{homogeneous} and \textit{ergodic}.
These technical conditions are made more explicit in Appendix \ref{appendix-probab-estimates}. Then the
corresponding random Schr\"{o}dinger operator acting in $\mathscr{H} := L^{2} (\mathbb{R}^d)$ is a
perturbation of the
kinetic-energy operator:
\begin{eqnarray}\label{Schrodinger-operator-inf}
h^{\omega} := - {\hbox{\small{$\frac{1}{2}\hskip 0.01cm$}}} \Delta \, \dotplus \, v^{\omega} ,
\end{eqnarray}
defined as a sum in the \textit{quadratic-forms} sense. The restriction to the box $\Lambda_{l}$,
is specified by the Dirichlet boundary conditions and for regular potentials one gets the self-adjoint
operator:
\begin{eqnarray}\label{Schrodinger-operator}
h_{l}^{\omega} := \left( - \shalf \Delta \, + \, v^{\omega} \right)_{D} = h_{l}^{0} \dotplus \, v^{\omega},
\end{eqnarray}
acting in $ \mathscr{H}_l$. We denote by $\{\phi_{i}^{\omega,l}, E_{i}^{\omega,l}\}_{i \geqslant 1}$ the
set of normalized eigenfunctions and corresponding eigenvalues  of $h_{l}$. Again, we order the eigenvalues
(counting the multiplicity) so
that $E_{1}^{\omega,l} \leqslant E_{2}^{\omega,l} \leqslant E_{3}^{\omega,l}\dots \,\,$. Note that the
\textit{non-negativity} of the random potential implies {that} $E_{1}^{\omega,l} > 0$. {So, for convenience we
assume also that in the thermodynamic limit \textit{almost surely} (a.s.) with
respect to the probability $\mathbb{P}$, the lowest edge of this random one-particle spectrum is:\\
(iii) a.s.-$\lim_{l \rightarrow \infty} E_{1}^{\omega,l} = 0$.

When no confusion arises, we shall \textit{omit} the explicit mention of $l$ and $\omega$ dependence.
Note that the non-negativity of the potential implies that:
\begin{eqnarray}\label{Positive-Potential}
&(a)& \, Q (h_{l}^{\omega})  \subset   Q (h_{l}^{0}), \quad Q\,\,\,\, \textrm{being the quadratic
form domain},\\
&(b)& \, (\varphi , h_{l}^{\omega} \varphi) \, \geqslant \, (\varphi , h_{l}^{0} \varphi), \,\,
\forall \varphi \in Q (h_{l}^{\omega}) .\nonumber
\end{eqnarray}

Now, we turn to the many-body problem. Let $\mathscr{F}_{l}:= \mathscr{F}_{l}(\mathscr{H}_{l})$ be the
symmetric Fock space constructed over $ \mathscr{H}_{l}$. Then $H_l:={\rm{d\Gamma}}(h_l^\omega)$ denotes
the second
quantization of the \textit{one-particle} Schr\"{o}dinger operator $h_{l}^{\omega}$ in $\mathscr{F}_{l}$.
Note that
the operator $H_l$ acting in $\mathscr{F}_{l}$ has the form:
\begin{equation}\label{Fock-one-part}
H_{l} =  \sum_{i \geqslant 1}  E_{i}^{\omega,l} \   a^{*}(\phi_{i}) a(\phi_{i}),
\end{equation}
where $a^{*}(\phi_{i}),  a(\phi_{i})$ are the creation and annihilation operators (satisfying the boson
\textit{Canonical Commutation Relations}) in the one-particle eigenstates
$\{\phi_{i}:= \phi_{i}^{\omega,l}\}_{i \,\geq 1}$
of $h_{l}^{\omega}$. Then, the grand-canonical Hamiltonian of the perfect Bose gas in a random external
potential is given by:
\begin{eqnarray}\label{multi.part.Hamiltonian}
H_{l} (\mu) \, := \, H_{l} - \mu N_{l} \, = \, \sum_{i \geqslant 1}  (E_{i}^{\omega,l} - \mu)\   N_{l}
(\phi_{i})
\end{eqnarray}
where $N_{l}(\phi_{i}) := a^{*}(\phi_{i}) a(\phi_{i})$ {is the operator for the number of particles in
the eigenstate $\phi_i$,
$N_{l} := \sum_{i} N_{l}(\phi_{i})$  is the operator for the total number of particles in $\Lambda_l$}
and $\mu$ is the
chemical potential. Note that $N_{l}$ can be expanded over \textit{any} basis in the space
$ \mathscr{H}_l $, and in
particular over the one defined by the {free one-particle kinetic-energy eigenstates $\{\psi_{k}^{l},
\varepsilon_k\}_k $.}

We denote by $\langle - \rangle_{l}$ the equilibrium state defined by the Hamiltonian $H_{l}(\mu)$ :
\begin{eqnarray*}
\langle A\rangle_{l}(\beta,\mu) :=
\frac{\textrm{Tr}_{ \mathscr{F}_{l}}\{\exp ( -\beta H_{l}(\mu) ) A\}}{\textrm{Tr}_{ \mathscr{F}_{l}}
\exp ( -\beta H_{l}(\mu) )}.
\end{eqnarray*}
For simplicity, we shall omit in the following the explicit mention of the dependence on the
thermodynamic parameters
$(\beta,\mu)$. Finally, we define the \textit{Thermodynamic Limit} (TL) as the limit, when
$l \rightarrow \infty$.
\section{Generalized BEC in one-particle \emph{random} eigenstates}\label{section-gBEC-random-eigenstates}
\setcounter{equation}{0}
\renewcommand{\theequation}{\arabic{section}.\arabic{equation}}

In this section {we consider the possibility of macroscopic occupation of the one-particle random
Schr\"{o}dinger
operator (\ref{Schrodinger-operator}) eigenstates $\{\phi_{i}\}_{i \geq 1}$}. Recall
that the
corresponding limiting IDS, $\nu(E)$, is defined as:
\begin{eqnarray}\label{density-states}
\nu (E) \, := \, \lim_{l \rightarrow \infty} \nu_{l}^{\omega} (E) \, =  \, \lim_{l \rightarrow
\infty} \frac{1}{V_{l}}
\sharp \{ i : E_{i}^{\omega, l} \leqslant E \}.
\end{eqnarray}
Although the finite-volume IDS, $ \nu_{l}^{\omega} (E)$, are random measures, one can check that for
homogeneous ergodic random potentials the limit (\ref{density-states}) has the property of
\textit{self-averaging} \cite{PF}.
This means that $\nu (E)$ is almost surely (a.s.) a \textit{non-random} measure. Let us define  a (random)
particle density
\emph{occupation measures} $m_{l}$ by:}
\begin{equation}\label{random-occup-measure}
m_{l} (A) \,\, := \,\, \frac{1}{V_{l}} \, \sum_{i: E_{i} \in A}\langle N_{l}(\phi_{i})\rangle_{l},
\quad A \subset \mathbb{R}.
\end{equation}
{Then using standard methods, one can prove that this sequence of measures has (a.s.) a non-random weak-limit
$m$,
see (\ref{expression-general-occupation-measure}) below. Moreover,} if the critical density
\begin{eqnarray}\label{critical-density}
\rho_{c} \, := \, \lim_{\mu \rightarrow 0} \int_{0}^{\infty} \frac{1}{e^{\beta (E- \mu)}-1} \nu (\textrm{d}E)
\end{eqnarray}
is finite, then one obtains a \emph{generalized} Bose-Einstein condensation ({g-BEC}) {in the sense that
this measure
$m$ has an atom at the bottom of the spectrum of the random Schr\"{o}dinger operator, which by (iii),
Section
\ref{model-notations}, is assumed to be at $0$:
\begin{eqnarray}\label{usual-BEC}
m (\{ 0 \}) \, = \, \lim_{\delta \downarrow 0} \lim_{l \rightarrow \infty} \sum_{i: E_{i} \leqslant
\delta}  \,
\frac{1}{V_{l}} \, \langle N_{l}(\phi_{i})\rangle_{l} \,\, = \,\, \left\{ \begin{array}{ll}
0 & \,\,\textrm{if} \,\, \overline{\rho} <  \rho_{c}\\
\overline{\rho} -\rho_{c} \,\,&\,\,  \textrm{if} \,\,  \overline{\rho} \geqslant  \rho_{c}
\end{array} \right.
\end{eqnarray}
where $\overline{\rho}$ denotes  a (fixed) mean density \cite{LPZ}, \cite{LZ}. Physically, this corresponds
to the macroscopic occupation of the set of eigenstates $\phi_{i}$ with energy close to the ground state
$\phi_{1}$. However,
we have to stress that BEC in this sense does \emph{not} necessarily imply a macroscopic occupation of
the ground state.
{In fact, the condensate can be spread over many (and even infinitely many) states.

These various situations correspond to classification of the g-BEC on the \textit{types} I, II and III,
introduced in
eighties by van den Berg-Lewis-Pul\'{e}, see e.g. \cite{vdBLP} or \cite{ZB}, \cite{PZ}. {The most striking
case is type
III when generalized BEC occurs in the sense of equation (\ref{usual-BEC}) even though \emph{none} of the
eigenstates
$\phi_{i}$ are macroscopically occupied.} The realization of different types depends on how the relative
gaps between the
eigenvalues $E_{i}$ at the bottom of the spectrum vanishes in the TL.} To our knowledge, analysis of this
behaviour in random system has only been realised in some particular cases, see \cite{LZ} for a
comprehensive presentation.
The concept of \textit{generalized} BEC is more stable then the standard one-mode BEC, since it depends on
the global
low-energy behaviour of the density of states, especially on its ability to make the critical density
(\ref{critical-density})
finite. We note also that, since the IDS (\ref{density-states}) is not random, the same it true for the
amount of the g-BEC
(\ref{usual-BEC}).

We can also obtain an explicit expression for the limiting measure $m$. Note that we have \textit{fixed} the
\textit{mean density} $\overline{\rho}$, which implies that we require the chemical potential {$\mu$ to
satisfy the equation:}
\begin{equation}\label{condition-mu-finite-volume}
\overline{\rho} \, = \, \langle N_{l}\rangle_{l}{(\beta,\mu)} =
\frac{1}{V_{l}} \, \sum_{i \geq 1} \frac{1}{e^{\beta (E_{i}^{\omega, l}- \mu)}-1} \ ,
\end{equation}
for any $l$. {Since the system is disordered, the unique solution $\mu_{l}^{\omega}:=
{\mu_{l}^{\omega}(\beta,\overline{\rho})}$ of this equation is a \textit{random} variable, which is a.s.
non-random
in the TL  \cite{LPZ}, \cite{LZ}.} In the rest of this paper we denote the non-random $\mu_{\infty} :=
\textrm{a.s.-}\lim_{l \rightarrow \infty}\mu_{l}^{\omega}$. By condition (iii), Section \ref{model-notations},
and by (\ref{eq-rho-mu}) it is a continuous function of $\overline{\rho} $ :
\begin{eqnarray}\label{mu-inf}
\mu_{\infty}(\beta,\overline{\rho}) \, = \, \left\{ \begin{array}{ll} \,
0 & \,\,\textrm{if} \,\, \overline{\rho} \geq  \rho_{c} \ , \\
\overline{\mu} < 0\,\,&\,\,  \textrm{if} \,\,  \overline{\rho} < \rho_{c} \ ,
\end{array} \right.
\end{eqnarray}
where $\overline{\mu}:= \overline{\mu}(\beta,\overline{\rho})$ is a (unique) solution of the equation:
\begin{eqnarray}\label{eq-rho-mu}
\overline{\rho} \, = \, \int_{0}^{\infty} \frac{1}{e^{\beta (E - \mu)}-1} \ \nu (\textrm{d}E) \ ,
\end{eqnarray}
for $\overline{\rho} \leq  \rho_{c}$.
\begin{remark}\label{random-chem-pot-values}
Note that $\mu_{\infty}$ is non-positive (\ref{mu-inf}), which is not true in general for the random
finite-volume solution $\mu_{l}^{\omega}$. Indeed, the only restriction we have is that
$\mu_{l}^{\omega} < E_{1}^{\omega, l}$, which is the well-known condition for the pressure of the perfect
Bose
gas to \textit{exist}. We return to this question in Section \ref{Sect-kinetic-occupation-measure} for the
case of
\textit{random} BEC in the free one-particle kinetic-energy operator eigenfunctions.
\end{remark}

We also recall that for (\ref{mu-inf}) the explicit expression of the weak limit for the general particle
density occupation measure is:
\begin{eqnarray}\label{expression-general-occupation-measure}
m(\ud E) \, = \, \left\{ \begin{array}{ll}
(\overline{\rho} - \rho_{c}) \delta_{0}(\ud E) \, + \, (e^{\beta E}-1)^{-1} \, \nu(\ud E) \
& \,\,\textrm{if} \,\, \overline{\rho} \geqslant  \rho_{c} \ , \\
(e^{\beta(E-\mu_{\infty})}-1)^{-1} \, \nu(\ud E)\,\,&\,\,  \textrm{if} \,\,  \overline{\rho} <  \rho_{c} \ .
\end{array} \right.
\end{eqnarray}

We end this section with a comment about the difference between the model of the perfect Bose gas
embedded into a
random potential and the \emph{free} Bose gas. In the latter case, one should consider the IDS of the
one-particle
kinetic-energy operator (\ref{Kinetic-energy-operator}), which is given by the \textit{Weyl formula}:
\begin{eqnarray}\label{density-states-kinetic}
\nu^{0} (E) \, = \,  C_{d}E^{d/2},
\end{eqnarray}
where is $C_{d}$ is a constant term depending only on the dimensionality $d$. It is known that for this IDS,
the
critical density (\ref{critical-density}) is finite only when $d > 2$, and hence the fact that BEC does
not occur for low
dimensions. On the other hand, a common feature of Schr\"{o}dinger operators with regular, stationary,
non-negative
ergodic random potentials is the so-called \emph{Lifshitz tails} behaviour of the IDS near the bottom of the
spectrum. {When the lower edge of the spectrum coincides with  $E=0$ (condition (iii)), this means roughly that
(see for example \cite{PF}):
\begin{equation}\label{doublelogarithmic}
\nu (E)\sim e^{-a/E^{d/2}}
\end{equation}
for small $E$ and $a>0$. Hence, the critical density (\ref{critical-density}) is finite in any dimension, 
and therefore enhances BEC in the sense of
(\ref{usual-BEC}) even for $d=1,2$.  This was shown in \cite{LPZ}, \cite{LZ}, where some specific examples of
one-dimensional \textit{Poisson disordered} systems exhibiting g-BEC in the sense of (\ref{usual-BEC}) were
studied. In this article we require only the following rigorous upper estimate:
\begin{eqnarray}\label{Lifshitz-tails-infinite-volume}
\lim_{E \rightarrow 0^{+}} \, (- E^{d/2}) \ln (\nu (E)) \geqslant a  >  0 \ ,
\end{eqnarray}
for some constant $a$. This can be proved (see \cite{KM}) under the technical conditions detailed in Appendix 
\ref{appendix-probab-estimates},
which are assumed throughout this paper.
In particular these conditions are satisfied in the case of Poisson random potentials with sufficiently fast
decay of the potential around each impurity.}

\section{Generalized BEC in one-particle \textit{kinetic} energy  eigenstates}
\label{Sect-kinetic-occupation-measure}
\setcounter{equation}{0}
\renewcommand{\theequation}{\arabic{section}.\arabic{equation}}
\subsection{Occupation measure for one-particle kinetic energy eigenstates}
\label{Sect-kinetic-occupation-measure-main-theorem}
{Similar to (\ref{random-occup-measure}), we introduce the sequence of particle occupation measure
$\tilde{m}_{l}$ for
kinetic energy eigenfunctions}
$\{\psi_{k}:=\psi_{k}^{l}\}_{k\in \Lambda_{l}^*}$:
\begin{equation}\label{expression-kinetic-occupation-measure}
\tilde{m}_{l} (A) \,\, := \,\, \frac{1}{V_{l}} \, \sum_{k: \varepsilon_k \in A}
\langle N_{l}(\psi_{k})\rangle_{l} \ , \quad A \subset \mathbb{R} \ ,
\end{equation}
but now in the \textit{random equilibrium states} $\langle -\rangle_{l}$ corresponding to the perfect
boson gas
with Hamiltonian (\ref{Fock-one-part}).

{Note that, contrary to the last section, the standard arguments used to prove the existence of a limiting
measure
in TL are not valid for (\ref{expression-kinetic-occupation-measure}), since the kinetic energy operator
(\ref{Kinetic-energy-operator}) and the random Schr\"{o}dinger operator (\ref{Schrodinger-operator})
\textit{do not commute}.}

{We remark also that even if we know that the measure $m$ (\ref{expression-general-occupation-measure})
has an atom at the edge of the spectrum (g-BEC), we cannot deduce that the limiting measure
$\tilde{m}$ (assuming that it exists) also manifests g-BEC in the free kinetic energy eigenstates $\psi_{k}$.}

Our motivation to study this problem is that in view of the well-known \textit{Bogoliubov approximation},
it is the later
that is required, see \cite{ZB}-\cite{KT}. Indeed, let the second-quantized form of the interaction
term is expressed in
terms of creation/annihilation operators in states $\psi_{k}^{l}$, eigenfunctions of kinetic-energy operator
(\ref{Kinetic-energy-operator}). Then the so-called first Bogoliubov approximation (Bogoliubov
\textit{ansatz}) assumes
that only terms involving creation and/or annihilation operators of particles in the ground state
$\psi_{1}^{l}$ are relevant.
The Bogoliubov theory is nontrivial if there is macroscopic occupation of this zero-mode kinetic-energy
operator ground-state.
Therefore, the g-BEC in the sense (\ref{usual-BEC}) is not sufficient to apply the Bogoliubov \textit{ansatz}.

Now we formulate the main result of this section. Let
\begin{eqnarray*}
\Omega_{(x,x')}^{T} := \{ \xi : \xi(0) = x, \, \xi(T) = x' \}
\end{eqnarray*}
be the set of continuous trajectories (paths) $\{\xi(s)\}_{s=0}^{T}$ in ${{\mathbb{R}}}^{d}$,
connecting the points $x$, $x'$, and let $w^{T}$ denote the normalized Wiener measure on this set.

{\begin{theorem}\label{existence-occupation-measure-and-BEC}
The sequence of measures $\tilde{m}_{l}$ converges a.s. in a weak sense to a non-random measure $\tilde{m}$,
which is given by:
\begin{eqnarray*}
\tilde{m}(\ud \varepsilon) \, = \,  \left\{ \begin{array}{ll}
               (\overline{\rho} - \rho_{c}) \delta_{0}(\ud \varepsilon)  \,
               + \, F(\varepsilon) \ud \varepsilon & \,\,\textrm{if} \,\, \overline{\rho} \geqslant
               \rho_{c}\\
              F(\varepsilon) \ud \varepsilon \,\,&\,\,  \textrm{if} \,\,  \overline{\rho} <  \rho_{c}
              \end{array} \right.
\end{eqnarray*}
with density $F(\varepsilon)$ defined by:
\begin{eqnarray*}
F(\varepsilon) \, = \, (2\varepsilon)^{d/2 - 1} \int_{S_{d}^{1}} \, \ud \sigma \, g(\sqrt{2\varepsilon} \,
{n}_{\sigma}) \
.\\
\end{eqnarray*}
Here, $S_{d}^{1}$ denotes the unit sphere in $\mathbb{R}^{d}$ centered at the origin, ${n}_{\sigma}$ the unit
outward drawn normal vector, and $\ud \sigma$ the surface measure of $S_{d}^{1}$. The function $g$ is
as follows
\begin{eqnarray}
\label{g-function}
g(k) \, = \, \frac{1}{(2\pi)^{d/2}} \, \int_{\mathbb{R}^{d}} \, \ud x \, e^{ikx} \, \sum_{n \geqslant 1}
e^{n\beta\mu_{\infty}}
\frac{e^{-\|x\|^2(1/2n\beta)}}{(2\pi n\beta)^{d/2}} \, \mathbb{E}_{\omega} \Big\{ \,
\int_{\Omega_{(0,x)}^{n\beta}} \,
w^{n\beta}(\ud\xi) \,
e^{-\int_{0}^{n\beta} \, \ud s \, v^{\omega} (\xi(s))}\Big\},
\end{eqnarray}
with expectation $\mathbb{E}_{\omega}$ on the probability space $(\Omega, \mathcal{F}, \mathbb{P})$.
\end{theorem}}

Note that since the measures $w^{n\beta}$ on $\Omega_{(0,x)}^{n\beta}$ are normalized, we recover
from (\ref{g-function}) the expression for the free Bose gas if we put $v^{\omega} = 0$.

\noindent Before proceeding with the proof, we give some comments about these results. \\
(a) First, the existence of a non-trivial limiting kinetic energy states occupation measure provides a
rigorous basis
for discussing the macroscopic occupation of the free Bose gas eigenstates.\\
(b) Moreover, both occupation measures (\ref{expression-general-occupation-measure}) and
(\ref{expression-kinetic-occupation-measure}) do not only exhibit simultaneously an atom at the bottom of the
spectrum, but these atoms have the \textit{same} non-random weights. It is quite surprising that the
generalized BEC
triggered by the Lifshitz tail in a low dimension {disordered} system produces the same value of the
generalized
BEC in the lowest free kinetic energy states.

\subsection{Proofs}
We start by expanding the measure $\tilde{m}$ in terms of {the random equilibrium mean-values of
occupation numbers in the corresponding eigenstates} $\phi_{i}$. Using the linearity {(respectively
conjugate linearity)} of the creation and annihilation operators one obtains:
\begin{eqnarray} \label{free-meas-A}
\tilde{m}_{l} (A)  &=&  \frac{1}{V_{l}} \, \sum_{k: \varepsilon_k \in A} \langle a^{*}(\psi_{k})
a(\psi_{k})\rangle_{l}\\
&=&  \frac{1}{V_{l}} \,\sum_{i,j} \sum_{k: \varepsilon_k \in A} \, \overline{(\phi_{i},\psi_{k})}
(\phi_{i},(\psi_{k}) \,
\langle a^{*}(\phi_{i}) a(\phi_{j})\rangle_{l}\nonumber \\
&=&  \frac{1}{V_{l}} \,\sum_{i} \sum_{k: \varepsilon_k \in A} \,  |(\phi_{i},\psi_{k})|^{2} \,
\langle a^{*}(\phi_{i}) a(\phi_{i})\rangle_{l}.\nonumber
\end{eqnarray}
In the last equality, we have used the fact that $\left[H_{l} (\mu) , N_{l}(\phi_{i}) \right] = 0$ for
all $i$, which implies that:
\begin{eqnarray*}
\langle a^{*}(\phi_{i}) a(\phi_{j})\rangle_{l} \, = \, 0 \quad \textrm{if} \ i \neq j .
\end{eqnarray*}
This is the analogue of the \textit{momentum conservation law} in the free Bose gas. {Although, it has a
different
physical meaning: the conservation of the particle number in each random eigenstate  $\phi_{i}$.}

{We first prove two important lemmas. In neither of them we shall assume that the sequence $\tilde{m}_{l}$
has a
weak limit, instead we consider only some convergent subsequence. Note that at least one such subsequence
always exists,
see \cite{F1}}.

The first result states that if there is condensation in the lowest random eigenstates $\{\phi_{i}\}_{i}$,
then there is
also condensation in the lowest kinetic-energy states $\{\psi_{k}\}_{k}$. Moreover, the amount of the latter
condensate
density has to be not \textit{less} than the former.
\begin{lemma}\label{lower-bound-kineticBEC}
Let $\{\tilde{m}_{l_{r}}\}_{r\geq 1}$ be a convergent subsequence. We denote by $\tilde{m}$ its (weak) limit.
Then:
\begin{eqnarray*}
\tilde{m} (\{0\}) \, \geqslant \, m(\{0\}) \, = \, \left\{ \begin{array}{ll}
\overline{\rho} - \rho_{c} & \ \ \textrm{if} \ \ \ \overline{\rho} \geqslant  \rho_{c}\\
0 \,\,& \ \  \textrm{if} \ \ \ \overline{\rho} <  \rho_{c} \ \ .
\end{array} \right.
\end{eqnarray*}
\end{lemma}
\textit{Proof}:
Let $\gamma>0$. Using the expansion of the functions $\psi_{k}$ in the basis $\{\phi_{i}\}_{i \geq 1}$ ,
we obtain:
\begin{eqnarray*}
\tilde{m} ([0,\gamma]) &=&  \lim_{r \rightarrow \infty} \,\frac{1}{V_{l_{r}}} \, \sum_{k:\varepsilon_{k}
\leqslant \gamma }\langle N_{l_{r}}(\psi_{k})\rangle_{l_{r}} \\
&=& \lim_{r \rightarrow \infty} \frac{1}{V_{l_{r}}} \, \sum_{k:\varepsilon_{k} \leqslant \gamma }
\sum_{i \geqslant 1} |(\phi_{i},\psi_{k})|^{2}  \,   \langle N_{l_{r}}(\phi_{i})\rangle_{l_{r}}\\
&\geqslant& \lim_{r \rightarrow \infty} \frac{1}{V_{l_{r}}} \, \sum_{k:\varepsilon_{k} \leqslant \gamma }
\sum_{i:E_{i} \leqslant \delta} |(\phi_{i},\psi_{k})|^{2}  \,   \langle N_{l_{r}}(\phi_{i})\rangle_{l_{r}}
\end{eqnarray*}
for any $\delta > 0$. The non-negativity of the random potential (\ref{Positive-Potential}) implies:
\begin{eqnarray*}
\sum_{k:\varepsilon_{k} > \gamma } |(\phi_{i},\psi_{k})|^{2}  \leqslant  \sum_{k:\varepsilon_{k} > \gamma }
\frac{\varepsilon_{k}}{\gamma}|(\phi_{i},\psi_{k})|^{2} \leqslant  \frac{1}{\gamma} \sum_{k \geqslant 1 }
\varepsilon_{k} |(\phi_{i}, \psi_{k})|^{2} = \frac{1}{\gamma} (\phi_{i}, {h_{l}^{0}} \phi_{i})
\leqslant \,  \frac{1}{\gamma}(\phi_{i}, h_{l}^{\omega} \phi_{i})  =  \frac{E_{i}^{\omega}}{\gamma} \ .
\end{eqnarray*}
We then obtain:
\begin{eqnarray*}
\tilde{m}([0,\gamma]) &\geqslant&  \lim_{r \rightarrow \infty} \,\frac{1}{V_{l_{r}}} \,  \sum_{i:  E_{i}
\leqslant \delta}  \,  \langle N_{l_{r}}(\phi_{i})\rangle_{l_{r}} \, (1 - \sum_{k:\varepsilon_{k} >
\gamma }|(\phi_{i},\psi_{k})|^{2})\\
&\geqslant&  \lim_{r \rightarrow \infty} \,\frac{1}{V_{l_{r}}} \,  \sum_{i:  E_{i} \leqslant \delta}  \,
\langle N_{l_{r}}(\phi_{i})\rangle_{l_{r}} \, (1- {E_{i}}/{\gamma})\\
&\geqslant&  \lim_{r \rightarrow \infty} (1-{\delta}/{\gamma}) \,\frac{1}{V_{l_{r}}} \,
\sum_{i:  E_{i} \leqslant \delta}  \,  \langle N_{l_{r}}(\phi_{i})\rangle_{l_{r}} \, = \,
(1-{\delta}/{\gamma}) \, m([0,\delta]) \, {\geq 0} \ .
\end{eqnarray*}
But $\delta$ is arbitrary, and the lemma follows by letting $\delta\to 0$.   \hfill $\square$

In the next lemma, we show that the measure $\tilde{m}$ (\ref{expression-kinetic-occupation-measure})
can have an atom only at zero kinetic energy.
\begin{lemma}\label{absolute-continuity-kinetic-occupation-measure}
Let $\{\tilde{m}_{l_{r}}\}_{r\geq 1}$ be a convergent subsequence, and $\tilde{m}$ be its (weak) limit.
Then, it is
absolutely continuous on $\mathbb{R}_{+}:=(0,\infty)$.
\end{lemma}
\textit{Proof}:
Let $A$ to be a Borel subset of $(0,\infty)$, {with Lebesgue measure $0$, and let $a$ be such that
$\inf A \, > \, a \, > \, 0 $. Then:}
\begin{eqnarray}\label{estim1}
\tilde{m}_{l_{r}}(A) &=& \frac{1}{V_{l_{r}}} \, \sum_{k: \varepsilon_k \in A} \,
\langle N_{l_{r}}(\psi_{k})\rangle_{l_{r}}\\
&=&  \frac{1}{V_{l_{r}}} \, \sum _{k: \varepsilon_k \in A} \, \sum_{i} \, |(\phi_i, \psi_k)|^{2} \,
\langle N_{l_{r}}(\phi_{i})\rangle_{l_{r}} \nonumber \\
&=& \frac{1}{V_{l_{r}}} \, \sum _{k: \varepsilon_k \in A} \, \sum_{i: E_i \leqslant \alpha} \,
|(\phi_i, \psi_k)|^{2} \,  \langle N_{l_{r}}(\phi_{i})\rangle_{l_{r}} \, + \,  \frac{1}{V_{l_{r}}} \,
\sum _{k: \varepsilon_k \in A} \, \sum_{i: E_i > \alpha} \, |(\phi_i, \psi_k)|^{2} \,
\langle N_{l_{r}}(\phi_{i})\rangle_{l_{r}} \nonumber
\end{eqnarray}
for some $\alpha > 0$. Next, we use (\ref{Positive-Potential}) to get the following estimate:
\begin{eqnarray*}
E_{i} \, = \, (\phi_{i}, h_{l}^{\omega} \phi_{i}) \, \geqslant \, (\phi_{i}, h_{l}^{0}\phi_{i}) \, =
\, \sum_{k} \, \varepsilon_k  |(\phi_{i}, \psi_{k})|^{2} \, \geqslant \, a \,
\sum _{k: \varepsilon_k \in A} \, |(\phi_{i}, \psi_{k})|^{2} \ .
\end{eqnarray*}
Since the the equilibrium value of the occupation numbers $\langle N_{l}(\phi_{i})\rangle_{l}=
\{e^{E_{i} -\mu}- 1\}^{-1}$
are decreasing with $i$, {the estimate (\ref{estim1}) implies:}
\begin{eqnarray}\label{estim2}
\tilde{m}_{l_{r}}(A) \, \leqslant \,  \frac{1}{V_{l_{r}}} \, \frac{1}{a} \sum_{i: E_{i} \leqslant \alpha}
\, E_{i}  \langle N_{l_{r}}(\phi_{i})\rangle_{l_{r}} \, + \,  \langle N_{l_{r}}(\phi_{i_{\alpha}})
\rangle_{l_{r}}
\, \frac{1}{V_{l_{r}}} \, \sum _{k: \varepsilon_k \in A} \, 1  \ ,
\end{eqnarray}
where $\phi_{i_{\alpha}}$ denotes the eigenstate of $h_{l}^{\omega}$ with the \textit{smallest} eigenvalue
 \textit{greater} than $\alpha$. {Using again the monotonicity and the finite-volume IDS
(\ref{density-states})
we can get an upper bound for the mean occupation number in the second term of (\ref{estim2}), since:}
\begin{eqnarray}\label{estim3}
\overline{\rho} \, = \,  \frac{1}{V_{l}} \, \sum_{i} \, \langle N_{l}(\phi_{i})\rangle_{l} \, \geqslant \,
\frac{1}{V_{l}} \, \sum_{i: E_i \leqslant \alpha} \, \langle N_{l}(\phi_{i})\rangle_{l} \, \geqslant \,
\langle N_{l}(\phi_{i_{\alpha}})\rangle_{l} \  \nu_{l}^{\omega}(\alpha) \ .
\end{eqnarray}
{Combining (\ref{estim2}) and (\ref{estim3}) we obtain:}
\begin{eqnarray}\label{estim4}
\tilde{m}_{l_{r}}(A) \, \leqslant \,\frac{\alpha \, \overline{\rho}}{a} \, +
\, \frac{\overline{\rho}}{\nu_{l_{r}}^{\omega}(\alpha)} \, \int_{A} \nu_{l_{r}}^{0} (\ud \varepsilon) \ .
\end{eqnarray}
Since the measure $ \nu^{0}$ (\ref{density-states-kinetic}) is absolutely continuous with respect to
the Lebesgue measure, and $\nu(\alpha)$ is strictly positive for any $\alpha > 0$ the limit
$r \rightarrow \infty$
in (\ref{estim4}) gives:
\begin{eqnarray*}
\tilde{m}(A) \, \leqslant \,\frac{\alpha \, \overline{\rho}}{a} \ ,
\end{eqnarray*}
{But $\alpha > 0$ can be chosen  arbitrary small and thus $\tilde{m}(A) = 0$. To finish the proof,} note that
any Borel subset of $(0, \infty)$ can be expressed as a countable union of disjoint subsets with non-zero
infimum.
Our arguments than can be applied to each of them.     \hfill $\square$
{\begin{remark}\label{remark-validity-lemmas}
Lemmas \ref{lower-bound-kineticBEC} and \ref{absolute-continuity-kinetic-occupation-measure} are
fairly general, since they require only the non-negativity of the potential and in the random case, ergodicity.
In particular they apply  to non-random (weak) external potentials, that we consider below, as well
as to some models of an interacting Bose gas, as long as the many-particles Hamiltonian still satisfies
the commutation relation $\left[H_{l} (\mu) , N_{l}(\phi_{i}) \right] = 0$. In particular this holds
in the case of
models with interactions, which are diagonal in the occupation number operators.
\end{remark}}

{Above we exploited the fact that the sequence $\{\tilde{m}_{l}\}_{l\geq 1}$ has at least one accumulation
point. However, to prove convergence, we need to make use of some particular and explicit features of the
perfect Bose gas, as well as more detailed information about the properties of the external (random)
potential. In particular, we shall need some estimates of the (random) finite volume integrated density
of states, see Lemma \ref{Lifshitz-tails}.}

{To this end let us denote by $P_{A}$ the orthogonal projection onto the subspace spanned by the
one-particle kinetic energy states $\psi_{k}$ with kinetic energy $\varepsilon(k)$ in the set $A$. Then using
the explicit
expression for the mean occupation $\langle a^{*}(\phi_{i}) a(\phi_{i})\rangle_{l}$ and (\ref{free-meas-A})
we obtain:}
\begin{eqnarray}\label{free-meas-A-b}
\tilde{m}_{l} (A)  =  \frac{1}{V_{l}} \, \textrm{Tr} \, P_{A} \, (e^{\beta(h_{l}^{\omega} -
\mu_{l})} - 1)^{-1}
= \sum_{n \geqslant 1} \,  \frac{1}{V_{l}} \, \textrm{Tr} \, P_{A} \, (e^{-n\beta(h_{l}^{\omega} -
\mu_{l})}) \ .
\end{eqnarray}
Now we split the measure (\ref{free-meas-A-b}) into two parts:
\begin{eqnarray}\label{occupation-measure-with-cutoff}
\tilde{m}_{l} \, &=& \, \tilde{m}_{l}^{(1)} \, + \, \tilde{m}_{l}^{(2)}\qquad \textrm{with}  \ , \\
\tilde{m}_{l}^{(1)} (A)\, &:=& \,  \sum_{n \geqslant 1} \,  \frac{1}{V_{l}} \, \textrm{Tr}
\, P_{A} \, (e^{-n\beta(h_{l}^{\omega} - \mu_{l})}) \, \mathbf{1} (\mu_{l} \leqslant 1/n)  \ , \nonumber\\
\tilde{m}_{l}^{(2)} (A)\, &:=& \,  \sum_{n \geqslant 1} \,  \frac{1}{V_{l}} \, \textrm{Tr}
\, P_{A} \, (e^{-n\beta(h_{l}^{\omega} - \mu_{l})}) \, \mathbf{1} (\mu_{l} > 1/n) \ . \nonumber
\end{eqnarray}
{Note that since the chemical potential satisfies equation (\ref{condition-mu-finite-volume}),
$\mu_l:=\mu_{l}^{\omega}$,
therefore the indicator functions $\mathbf{1} (\mu_{l} \leqslant 1/n)$ and $\mathbf{1} (\mu_{l} > 1/n)$
split the range of $n$ into the sums (\ref{occupation-measure-with-cutoff}) in a random and
volume-dependent way.}

We start with the proof of existence of a weak limit of the sequence of random measures
$\tilde{m}_{l}^{(1)} $:
\begin{theorem}\label{Laplace-transform-occupation-measure}
Let random potential $v^{\omega}$ satisfy the assumptions {\rm{(i)-(iii)}} of Section 2. Then for any
$d \geqslant 1$, the sequence of Laplace transforms of the measures $\tilde{m}_{l}^{(1)}$:
\begin{equation}\label{Lap-fin}
f_{l}(t; \beta, \mu_{l}) \,\,\, := \,\,\,  \int_{\mathbb{R}} \, \tilde{m}_{l}^{(1)}(\ud \varepsilon)
\ e^{-t\varepsilon}
\end{equation}
converges for any $t > 0$ to a (non-random) limit $f(t; \beta, \mu_{\infty})$ , which is given by:
\begin{equation}\label{Lap-infin}
f(t; \beta, \mu_{\infty}) \, = \, \sum_{n \geqslant 1} e^{n\beta\mu_{\infty}} \, \int_{\mathbb{R}^{d}}
\ud x \, \frac{e^{-\|x\|^2(1/2n\beta+1/2t)}}{(4\pi^2tn\beta)^{d/2}} \, \mathbb{E}_{{\omega}}
\Big\{\int_{\Omega_{(0,x)}^{n\beta}} \, w^{n\beta}(\ud \xi) \, e^{-\int_{0}^{n\beta} \,
\ud s \, v^{\omega}
(\xi(s))}\Big\} \ .
\end{equation}
Here $\mathbb{E}_{{\omega}}$ denotes the expectation with respect to realizations (configurations)
$\omega$ of the random
potential. Note that the sum on the right-hand side converges for {all (non-random) $\mu_{\infty}\leq 0 $},
including $0$, which corresponds to the case $\overline{\rho} \geqslant \rho_{c}$.
\end{theorem}
\textbf{Proof}: By definition of $P_{A}$ the Laplace transformation (\ref{Lap-fin}) can be written as:
\begin{eqnarray}\label{series-1}
f_{l}(t; \beta, \mu_{l}) = \sum_{n \geqslant 1} \,  \frac{1}{V_{l}} \, \textrm{Tr} \, e^{-t \, h_{l}^{0}} \,
(e^{-n\beta(h_{l}^{\omega} - \mu_{l})}) \, \mathbf{1} (\mu_{l} \leqslant 1/n) \ .
\end{eqnarray}
Now we have to show the uniform convergence of the sum over $n$ to be able to take the
term by term limit with respect to $l$. Since for any bounded operator $A$ and for any trace-class
non-negative operator $B$ one has $\textrm{Tr} AB \leqslant \|A\|\, \textrm{Tr} B $, we get
\begin{eqnarray}\label{definition-coeff-aln}
a_{l}(n) &:=& \frac{1}{V_{l}} \textrm{Tr} \, e^{-t \, h_{l}^{0}} \, e^{ -n \beta(h_{l}^{\omega} - \mu_{l})}
\mathbf{1}(\mu_{l} \leqslant 1/n)\\
&\leqslant& \frac{1}{V_{l}} \textrm{Tr} \, e^{ - n \beta(h_{l}^{\omega} - \mu_{l})} \mathbf{1}(\mu_{l}
\leqslant 1/n) \ .
\nonumber
\end{eqnarray}
For $\overline{\rho} < \rho_{c}$, the uniform convergence in (\ref{Lap-infin}) is immediate. Indeed, for
$l$ large enough, the chemical potential satisfies $\mu_{l} < \mu_{\infty} / 2 < 0$, which {by
(\ref{density-states}) provides  the following  a.s. estimate for
(\ref{definition-coeff-aln}):
\begin{eqnarray}\label{upper-bound-a_l(n)-rho-less-rho_c}
a_{l}(n) \, \leqslant \, e^{n \beta \mu_{\infty} / 2} \int_{[0,\infty)} {\nu_{l}^{\omega}(\ud E)} \
e^{-\beta E}  \,
\leqslant \, K_{1} \, e^{n \beta \mu_{\infty} / 2},
\end{eqnarray}
with some constant $K_{1}$.

However, for the case $\overline{\rho} \geqslant \rho_{c}$, this approach does not
work, since, in fact, for any finite $l$ the solutions $\mu_{l}= {\mu_{l}^{\omega}}$ of equation
(\ref{condition-mu-finite-volume}) {could} be \textit{positive} with some probability, event though by
condition (iii)
(see Section 2) it has to \textit{vanish} a.s. in the TL. We use, therefore, the bound:
\begin{eqnarray*}
a_{l}(n) &\leqslant&  a^{1}_{l}(n) \, + \, a^{2}_{l}(n) \ , \\
a^{1}_{l}(n) &:=& \frac{1}{V_{l}} \ e^{\beta} \ \sum_{\{i: \, E_{i}^{\omega,l} \leqslant 1/n^{1-\eta}\}}
e^{-n \beta E_{i}^{\omega,l}} \ , \\
a^{2}_{l}(n) &:=& \frac{1}{V_{l}} \ e^{\beta}\ \sum_{\{i: \, E_{i}^{\omega,l} > 1/n^{1-\eta}\}}
e^{-n \beta E_{i}^{\omega,l}}\nonumber \ ,
\end{eqnarray*}
which follows, for some $0< \eta < 1$, from the constraint $\mu_{l} n \leqslant 1$ due to the indicator
function in
(\ref{definition-coeff-aln}).
Then the first term is bounded from above by:
\begin{eqnarray*}
a^{1}_{l}(n)  \leqslant e^{\beta} \, \nu_{l}^{\omega} \, (n^{ \eta-1}) \ .
\end{eqnarray*}

On the other hand, by Theorem \ref{Lifshitz-tails} (\textit{finite-volume} Lifshitz tails), for
$\alpha > 0$ and  $0 < \gamma < d/2$, {there exists a subset $\widetilde{\Omega}\subset\Omega$
of full measure, $\mathbb{P}(\widetilde{\Omega}) =1$, such that for any $\omega\in \widetilde{\Omega}$
there exists a positive finite energy $\mathcal{E}(\omega):= \mathcal{E}_{\alpha,\gamma}(\omega) > 0$
for which one obtains:
\begin{eqnarray*}
\nu_{l}^{\omega} (E) \, \leqslant \, e^{-\alpha/ E^{\gamma}} \ ,
\end{eqnarray*}
for all $E < \mathcal{E}(\omega)$ .}
Therefore, for any configuration $\omega\in \widetilde{\Omega}$ (i.e. almost surely) we have
the {\textit{volume independent} estimate for all $n> \mathcal{N}(\omega):=\mathcal{E}
(\omega)^{1/(\eta-1)}$}:
\begin{eqnarray}\label{upper-bound-a^1_n}
a^{1}_{l}(n) \, \leqslant  e^{\beta} \, e^{-\alpha n^{(1-\eta)\gamma}} \ .
\end{eqnarray}

To estimate the coefficients $a^{2}_{l}(n)$ from above , we use the upper bound:
\begin{eqnarray*}
a^{2}_{l}(n) \, &\leqslant& \,  \, \int_{\left[1/n^{1-\eta}, \infty\right)} \, \nu_{l}^{\omega}
(\ud E) \ e^{-n\beta E}
\, \leqslant \,  e^{-\beta n^{\eta}/2} \int_{\left[1/n^{1-\eta},
\infty\right)} \, \nu_{l}^{\omega}(\ud E) \ e^{-n\beta E/2} \\ &\leqslant&
e^{-\beta n^{\eta}/2} \int_{\left[0, \infty\right)} \, \nu_{l}^{\omega}(\ud E) \ e^{-\beta E/2}  \ \ .
\end{eqnarray*}
Then for some $K_{2}> 0$ independent of $l$ we obtain:
\begin{eqnarray}\label{upper-bound-a^2_n}
a^{2}_{l}(n) \,\leqslant \, K_{2} e^{-\beta n^{\eta}/2} \ .
\end{eqnarray}

Therefore, by (\ref{upper-bound-a_l(n)-rho-less-rho_c}) in the case
$\overline{\rho} < \rho_{c}$, and by (\ref{upper-bound-a^1_n}), (\ref{upper-bound-a^2_n}) for
$\overline{\rho} \geq \rho_{c}$, we find that there exists a sequence $a(n)$ (independent of $l$) such that:
\begin{eqnarray}\label{uniform-convergence-ml1}
a_{l}(n) \leqslant a(n) \quad \textrm{and} \quad \sum_{n \geqslant 1} a(n) < \infty  \ .
\end{eqnarray}
Thus, the series {(\ref{series-1})} is uniformly convergent in $l$, and one can exchange sum and the limit:
\begin{eqnarray*}
\lim_{l \rightarrow \infty}f_l(t) =  \lim_{l \rightarrow \infty} \sum_{n = 0}^{\infty} {a_{l}(n) =
\sum_{n = 0}^{\infty}
\lim_{l \rightarrow \infty} a_{l}(n)} \ .
\end{eqnarray*}

The rest of the proof is largely inspired by the paper \cite{LPZ}. Using the Feynman-Kac representation,
we obtain the
following limit:
\begin{eqnarray}\label{Pastur-trick-original-version}
\lim_{l \rightarrow \infty} a_{l} (n) &=&\lim_{l \rightarrow \infty} \, \frac{1}{V_{l}} \textrm{Tr} \,
e^{-t \, h_{l}^{0}} \, e^{ -n \beta(h_{l}^{\omega} - \mu_{l})}  \mathbf{1}(\mu_{l} \leqslant 1/n)\\
&=& \lim_{l \rightarrow \infty} \, \frac{1}{V_{l}} \, \int_{\Lambda_{l}} \, \int_{\Lambda_{l}}
\ud x \,\ud x' \, e^{-t \, h_{l}^{0}} (x, x') \, e^{ -n \beta(h_{l}^{\omega} - \mu_{l})} (x',x)\nonumber\\
&=& e^{n\beta\mu_{\infty}} \lim_{l \rightarrow \infty} \, \frac{1}{V_{l}} \, \int_{\Lambda_{l}} \,
\int_{\Lambda_{l}} \ud x \, \ud x' \, \frac{e^{-\|x-x'\|^2(1/2n\beta+1/2t)}}{(4\pi^2tn\beta)^{d/2}} \times
\nonumber\\
&\,& \times \int_{\Omega_{(x',x)}^{n\beta}} \, w^{n\beta}(\ud \xi) \, e^{-\int_{0}^{n\beta} \,
\ud s \,v^{\omega}
(\xi(s))} \, \chi_{\Lambda_{l},n\beta} (\xi)\int_{\Omega_{(x,x')}^{t}} \, w^{t}(\ud \xi') \,
\chi_{\Lambda_{l},t} (\xi'),\nonumber
\end{eqnarray}
where we denote by $\chi_{\Lambda_{l}, T} (\xi)$ the characteristic function of paths $\xi$ such that
$\xi(t) \in \Lambda_{l}$ for all $0 < t < T$. Using Lemma \ref{estimate-finite-volume-errors}, we can
eliminate
these restrictions, and also extend one spatial integration over the whole space:
\begin{eqnarray}\label{Pastur-trick-Rd-no-paths-restriction}
&&\lim_{l \rightarrow \infty} \, \frac{1}{V_{l}} \textrm{Tr} \, e^{-t \, h_{l}^{0}} \,
e^{ -n \beta(h_{l}^{\omega} - \mu_{l})} \ = \\
&& e^{n\beta\mu_{\infty}} \lim_{l \rightarrow \infty} \, \int_{\mathbb{R}^{d}} \ud x\,
\frac{1}{V_{l}}\int_{\Lambda_{l}}  \ud x' \, \frac{e^{-\|x-x'\|^2(1/2n\beta+1/2t)}}{(4\pi^2tn\beta)^{d/2}}
\int_{\Omega_{(x',x)}^{n\beta}} \, w^{n\beta}(\ud \xi) \, e^{-\int_{0}^{n\beta} \, \ud s \, v^{\omega}
(\xi(s))} \ . \nonumber
\end{eqnarray}
Now, by the \textit{ergodic} theorem, we obtain:
\begin{eqnarray}\label{lim a}
&\,&\lim_{l \rightarrow \infty} a_{l} (n)= \lim_{l \rightarrow \infty} \, \frac{1}{V_{l}} \textrm{Tr} \, 
e^{-t \, h_{l}^{0}} \,
e^{ -n \beta(h_{l}^{\omega} - \mu_{l})}\\
&=& e^{n\beta\mu_{\infty}} \lim_{l \rightarrow \infty} \frac{1}{V_{l}}\int_{\Lambda_{l}}  \ud x'
\left\{\int_{\mathbb{R}^{d}} \ud x \frac{e^{-\|x-x'\|^2(1/2n\beta+1/2t)}}{(4\pi^2tn\beta)^{d/2}}
\int_{\Omega_{(x',x)}^{n\beta}} \, w^{n\beta}(\ud \xi) \, e^{-\int_{0}^{n\beta} \, \ud s \, v^{\omega}
(\xi(s))}\right\} \nonumber \\
&=& e^{n\beta\mu_{\infty}} \mathbb{E}_{\omega} \Big\{\int_{\mathbb{R}^{d}} \ud x
\frac{e^{-\|x\|^2(1/2n\beta+1/2t)}}
{(4\pi^2tn\beta)^{d/2}}\int_{\Omega_{(0,x)}^{n\beta}} \, w^{n\beta}(\ud \xi) \, e^{-\int_{0}^{n\beta}
\, \ud s \, v^{\omega} (\xi(s))}\Big\} \ . \nonumber
\end{eqnarray}
We then get the explicit expression for the limiting Laplace transform:
\begin{eqnarray*}
f(t; \beta, \mu_{\infty}) \, = \, \sum_{n \geqslant 1} e^{n\beta\mu_{\infty}} \, \int_{\mathbb{R}^{d}}
\ud x \frac{e^{-\|x\|^2(1/2n\beta+1/2t)}}{(4\pi^2tn\beta)^{d/2}} \, \mathbb{E}_{\omega}
\Big\{ \int_{\Omega_{(0,x)}^{n\beta}} \, w^{n\beta}(\ud \xi) \, e^{-\int_{0}^{n\beta} \,
\ud s \, v^{\omega} (\xi(s))}\Big\} \ ,
\end{eqnarray*}
which finishes the proof. \hfill $\square$
\begin{corollary}\label{existence-momentum-occupation-measure}
For any $\overline{\rho}$ the sequence of random measures $\tilde{m}_{l}^{(1)}$ converges a.s. in the
weak sense to a bounded, absolutely continuous non-random measure $\tilde{m}^{(1)}$, with density
$F(E)$ given by
\begin{eqnarray*}
F(\varepsilon) \, = \, (2\varepsilon)^{d/2 - 1} \int_{S_{d}^{1}} \, \ud \sigma \, g(\sqrt{2\varepsilon}
\, {n}_{\sigma}) \, .\\
\end{eqnarray*}
Here, $S_{d}^{1}$ denotes the unit sphere in $\mathbb{R}^{d}$, ${n}_{\sigma}$ the outward
drawn normal unit vector, $\ud \sigma$ the surface measure on $S_{d}^{1}$ and the function $g$ has the form
\begin{eqnarray}\label{g}
g(k) =  \frac{1}{(2\pi)^{d/2}} \, \int_{\mathbb{R}^{d}} \, \ud x \, e^{ik \cdot x} \,
\sum_{n \geqslant 1} e^{n\beta\mu} \
\frac{e^{-\|x\|^{2}/2n\beta}}{(2\pi n\beta)^{d/2}} \ \mathbb{E}_{\omega} \Big\{
\int_{\Omega_{(0,x)}^{n\beta}} \,
w^{n\beta}(\ud \xi) \, e^{-\int_{0}^{n\beta} \, \ud s \, v^{\omega} (\xi(s))}\Big\}.
\end{eqnarray}
\end{corollary}
\textbf{Proof}: By Theorem \ref{Laplace-transform-occupation-measure}, the existence of the weak limit
$\tilde{m}^{(1)}$
follows from the existence of the limiting Laplace transform. Moreover, we have the following explicit
expression: {
\begin{eqnarray*}
{\int_{\mathbb{R}} \tilde{m}^{(1)}(\ud \varepsilon) \ e^{-t\varepsilon}}  &=& \int_{\mathbb{R}^{d}} \ud x \
\frac{e^{-\|x\|^2/2t}}{(2\pi t)^{d/2}} \ \sum_{n \geqslant 1} e^{n\beta\mu} \
\frac{e^{-\|x\|^2/2n\beta}}{(2\pi n\beta)^{d/2}} \ \mathbb{E}_{\omega}
\Big\{ \int_{\Omega_{(0,x)}^{n\beta}} \,
w^{n\beta}(\ud \xi) \, e^{-\int_{0}^{n \beta} \ \ud s \, v^{\omega} (\xi(s))}\Big\} \\
&=&  \int_{[0, \infty)} \ud r\, e^{-t\|r\|^2/2} \, r^{d-1} \,  \int_{S_{d}^{1}} \ \ud \sigma \,
g(r \, {n}_{\sigma}) \\
&=&  \int_{[0, \infty)} \ud \varepsilon\, e^{-t\varepsilon} \, (2\varepsilon)^{d/2-1} \,
\int_{S_{d}^{1}} \, \ud \sigma \, g(\sqrt{2\varepsilon} \, {n}_{\sigma}) \ ,
\end{eqnarray*}
which proves the corollary . \hfill $\square$}

\begin{corollary}\label{weight-measure-m1}
The measure $\tilde{m}^{(1)}$ satisfies the following property:
\begin{eqnarray*}
\int_{[0,\infty)}  \, \tilde{m}^{(1)} ( \ud \varepsilon) \,\, = \,\,  \left\{ \begin{array}{ll}
\overline{\rho} & \,\,\textrm{if} \,\, \overline{\rho} <  \rho_{c}\\
\rho_{c} \,\,&\,\,  \textrm{if} \,\,  \overline{\rho} \geqslant  \rho_{c}
\end{array} \right.
\end{eqnarray*}
\end{corollary}
\textbf{Proof}: {By virtue of (\ref{series-1}) we have:
\begin{eqnarray*}
\int_{[0,\infty)}  \, \tilde{m}^{(1)} ( \ud \varepsilon) = f(0; \beta, \mu_{\infty}) = \lim_{l \rightarrow
\infty} \sum_{n \geqslant 1} \  \frac{1}{V_{l}} \ \textrm{Tr} \, e^{-n \beta(h_{l}^{\omega} - \mu_{l})}
\, \mathbf{1} (\mu_l \leqslant 1/n) \ .
\end{eqnarray*}}
Note that by uniformity of convergence of the sum, see (\ref{upper-bound-a^1_n}), (\ref{upper-bound-a^2_n}),
we can take the limit term by term (for any value of $\overline{\rho}$), and then:
\begin{eqnarray*}
&&\int_{[0,\infty)}  \, \tilde{m}^{(1)} ( \ud \varepsilon) = \sum_{n \geqslant 1} \,
\lim_{l \rightarrow \infty} \
\frac{1}{V_{l}} \ \textrm{Tr} \, e^{-n\beta(h_{l}^{\omega} - \mu_{l})} = \\
&&\sum_{n \geqslant 1}  \,\int_{[0,\infty)} \, \nu(\ud E) \, e^{-n\beta(E - \mu_{\infty})}
= \int_{[0,\infty)} \, \nu(\ud E) \, (e^{\beta(E - \mu_{\infty})} -1)^{-1} \ ,
\end{eqnarray*}
where we use Fubini's theorem for the last step.  \hfill $\square$

We are now ready for the proof of the main result of this section:

\noindent \textbf{Proof of Theorem \ref{existence-occupation-measure-and-BEC}}:
We first treat the case $\overline{\rho} < \rho_{c}$. In this situation, the measure $\tilde{m}_{l}^{(2)}$
is equal
to $0$ for $l$ large enough, see (\ref{occupation-measure-with-cutoff}), since the solution
$\lim_{l\rightarrow\infty}\mu_{l}^{\omega}$ of the equation
(\ref{condition-mu-finite-volume}) in the TL is a.s. strictly
negative.} Thus, the total occupation measure $\tilde{m}_{l}$ is reduced to $\tilde{m}^{(1)}_{l}$
and the Theorem follows from Corollary \ref{existence-momentum-occupation-measure}.

Now, consider the case $\overline{\rho} \geqslant \rho_{c}$. Choose a subsequence $l_{r}$ such that
the total
kinetic-energy states occupation measures $\tilde{m}_{l_{r}}$ converge weakly and a.s., and let the measure
$\tilde{m}$ be its limit.
By Corollary \ref{existence-momentum-occupation-measure}, all subsequences of measures
$\tilde{m}_{l_{r}}^{(1)}$
converge to the limiting measure $\tilde{m}^{(1)}$. Therefore,
by (\ref{occupation-measure-with-cutoff}), we obtain the weak a.s. convergence:
\begin{eqnarray*}
\lim_{r \rightarrow \infty} \, \tilde{m}_{l_{r}}^{(2)} \, =: \, \tilde{m}^{(2)} \ .
\end{eqnarray*}
By Lemma \ref{absolute-continuity-kinetic-occupation-measure}, we know that the measure $\tilde{m}$ is
absolutely continuous on $(0,\infty)$, and by Corollary \ref{existence-momentum-occupation-measure} that
$\tilde{m}^{(1)}$ is absolutely continuous on $[0,\infty)$. Therefore we get:
\begin{eqnarray*}
\tilde{m}^{\textrm{a.c.}} = \tilde{m}^{(1)}  +  \tilde{m}^{(2)\textrm{a.c.}} \ ,
\end{eqnarray*}
where $a.c.$ denotes the \textit{absolute continuous} components.

By definition of the total measure (\ref{occupation-measure-with-cutoff}),
$\tilde{m}([0,\infty))=\overline{\rho}$ and by Lemma
\ref{lower-bound-kineticBEC}, $\tilde{m}(\{0\})\geq\overline{\rho}-\rho_c$. Thus,
$\tilde{m}((0,\infty))\leq\rho_c$ and by Corollary \ref{weight-measure-m1}, we can then deduce
that the measure $\tilde{m}^{(2)}$ has no absolutely
continuous component and therefore consists at most of an atom at $\varepsilon=0$. Consequently, the
full measure $\tilde{m}$ can  be expressed as:
\begin{eqnarray*}
\tilde{m} \, = \, \tilde{m}^{\textrm{a.c.}} \, + \, b \delta_{0} \, = \, \tilde{m}^{(1)} \, +
\, b \delta_{0} \ ,
\end{eqnarray*}
and since by Corollary \ref{weight-measure-m1}
\begin{eqnarray*}\label{value-of-atom}
b  \, = \,  \overline{\rho} \, - \, \int_{\mathbb{R}_{+}} \, \tilde{m}_{l_{r}}^{\textrm{a.c.}}
(\ud \varepsilon) \,=
\, \overline{\rho} \, - \, \int_{\mathbb{R}_{+}} \, \tilde{m}_{l_{r}}^{(1)} (\ud \varepsilon)
\, = \, \overline{\rho} - \rho_{c}
\end{eqnarray*}
for the converging subsequence $\tilde{m}_{l_{r}}$, we have:
\begin{eqnarray*}
\lim_{l_{r} \rightarrow \infty} \, \tilde{m}_{l_{r}} \, = \, \tilde{m}^{(1)} \, + \, (\overline{\rho} -
\rho_{c}) \delta_{0} \ .
\end{eqnarray*}
By (\ref{value-of-atom}) and Corollary \ref{existence-momentum-occupation-measure}, this limit is
independent of the subsequence. Then, the limit of any convergent subsequence is the same, and  therefore,
using Feller's selection theorem, see \cite{F2}, the total kinetic states occupation measures
$\tilde{m}_{l}$ converge weakly to this limit. \hfill $\square$

\section{Finite volume Lifshitz tails}\label{Sect-Lifshitz-tails}
\setcounter{equation}{0}
\renewcommand{\theequation}{\arabic{section}.\arabic{equation}}
In this section, we give the proof of one important building block of our analysis,
Theorem \ref{Lifshitz-tails} about the \textit{finite-volume} Lifshitz tails.
Recall that this behaviour is a well-known feature of disordered systems, essentially meaning that for
 Shr\"{o}dinger operators which are semi-bounded from below, there are exponentially few eigenstates with energy close to
the bottom of the spectrum. To our knowledge, however, this is {always} shown only in the \textit{infinite-volume}
limit, see e.g. \cite{PF}. Here, we derive a \textit{finite-volume} estimate for the density of states, uniformly in {$l$},
though it could be trivial for small volumes. As one would expect our result is weaker than the asymptotic
one, in the sense that we prove it for Lifshitz exponent smaller than the limiting one.
\begin{theorem}\label{Lifshitz-tails}
Let the random potential $v^{\omega}$ satisfy the assumptions {\rm{(i)-(iii)}} of Section 2. Then
for any $\alpha > 0$ and $0 < \gamma < d/2$, {there exists a set $\widetilde{\Omega}\subset\Omega$
of full measure, $\mathbb{P}(\widetilde{\Omega}) =1$, such that for any configuration $\omega\in \widetilde{\Omega}$
one can find a positive finite energy $\mathcal{E}(\omega):= \mathcal{E}_{\alpha,\gamma}(\omega)$, for which one has
the estimate:
\begin{eqnarray*}
\nu_{l}^{\omega} (E) \, \leqslant \, e^{-\alpha /E^{\gamma}} \ \
\end{eqnarray*}
for all $E < \mathcal{E}(\omega)$ and for all $l$. }
\end{theorem}
For the proof, we first need a result from \cite{KM}.
\begin{lemma}\label{upper-bound-eigenvalue1}
By assumption {\rm{(ii)}} {\rm{(}}Section 2{\rm{)}} one has,
\begin{eqnarray*}
p \, = \, \mathbb{P} \Big\{\omega:  v ^{\omega}(0) = 0\Big\} \, < \, 1.
\end{eqnarray*}
Let $\alpha > p/(1-p)$, $B = \pi / (1+\alpha)$, and $E_{1}^{\omega,\, l, N}:= E_{1}^{\omega,N}$
be the
first eigenvalue of the random Schr\"{o}dinger operator (\ref{Schrodinger-operator}) with Neumann
{\rm{(}}instead of Dirichlet {\rm{)}} boundary conditions. Then, for $l$ large enough, there exists an
independent of $l$ constant $A = A(\alpha)$, such that
\begin{eqnarray}\label{estim-lemma5.1}
\mathbb{P} \left\{\omega: E_{1}^{\omega, N} < {B}/{l^{2}}\right\} \, < \, e^{-A V_{l}} \ .
\end{eqnarray}
\end{lemma}

Detailed conditions on the random potential and a sketch of the proof of this lemma are given in
Appendix \ref{appendix-probab-estimates}. Now we use Lemma \ref{upper-bound-eigenvalue1} to prove
the following result:
\begin{lemma}\label{lemma-proof-Lifshitz-tails}
Assume that the random potential satisfies the assumptions of Lemma \ref{upper-bound-eigenvalue1}.
Then for any $\alpha > 0$ and $0 < \gamma < d/2$,
\begin{eqnarray*}
\sum_{n \geqslant 1} \mathbb{P}\left\{ \sharp \Big\{i: E_{i}^{\omega,l} < {1}/{n}\Big\} >
V_{l} \, e^{-\alpha n^{\gamma}} \ , \ \textrm{\emph{for some}} \,\, l\geqslant 1 \right\} \, < \, \infty \ .
\end{eqnarray*}
\end{lemma}
\textbf{Proof}: Notice  that
\begin{eqnarray}\label{Proba-S}
\sum_{n \geqslant 1} \mathbb{P} \left\{ \sharp \Big\{i: E_{i}^{\omega,l} <
{1}/{n}\Big\} > V_{l} \, e^{-\alpha n^{\gamma}} \ ,  \ \textrm{for some} \,\,l\geqslant 1 \right\} \, =
\, \sum_{n \geqslant 1} \mathbb{P} \left\{ \bigcup_{l\geqslant 1} \, S_{l}^{n} \right\}  \ ,
\end{eqnarray}
where $S_{l}^{n}$ is the set
\begin{eqnarray}
\quad S_{l}^{n} := \Bigg\{ \omega: \, \sharp \Big\{i: E_{i}^{\omega,l} <
\frac{1}{n}\Big\} > V_{l} \, e^{-\alpha n^{\gamma}} \Bigg\}\nonumber \ .
\end{eqnarray}
The sum in the right-hand side of (\ref{Proba-S}) does not provide a very useful upper bound, since the
sets $S_{l}^{n}$ are highly overlapping. We thus need to define a new refined family of sets to avoid this
difficulty.

To this end we let $[a]_+$ be the smallest integer $\geqslant a$, and we define the
family of sets:
\begin{eqnarray*}
V_{k}^{n} := \Bigg\{ \omega: \, \sharp \Big\{i: E_{i}^{\omega,
\left[ (k e^{\alpha n^{\gamma}})^{1/d} \right]_{+}} <
\frac{1}{n}\Big\} \geqslant k \Bigg\} \ .
\end{eqnarray*}
Let $k := \left[ V_{l}e^{-\alpha n^{\gamma}} \right]_{+}$. Since $V_{l}= l^d$, this implies that
$h_{l}^{\omega} \geqslant h_{\left[(k e^{\alpha n^{\gamma}})^{1/d} \right]_{+}}^{\omega}$, and therefore:
\begin{eqnarray*}
\sharp \Big\{i: E_{i}^{\omega,\left[ (k e^{\alpha n^{\gamma}})^{1/d} \right]_{+}} < \frac{1}{n}\Big\}
\, \geqslant \,  \sharp \Big\{i: E_{i}^{\omega,l} < \frac{1}{n}\Big\} \ .
\end{eqnarray*}
If now $\omega \in S_{l}^{n}$, then by the definition of $k$ we obtain:
\begin{eqnarray*}
\sharp \Big\{i: E_{i}^{\omega,l} < \frac{1}{n}\Big\}  \, \geqslant \,  k  \  \ ,
\end{eqnarray*}
since the left-hand side is itself an integer. Thus, $S_{l}^{n} \subset V_{k}^{n}$ and:
\begin{eqnarray}\label{Proba-V}
\mathbb{P} \Big( \bigcup_{l\geqslant 1} \, S_{l}^{n} \Big) \, \leqslant
\, \mathbb{P} \Big( \bigcup_{k \geqslant 1} \, V_{k}^{n} \Big) \ .
\end{eqnarray}
We define also the sets:
\begin{eqnarray}\label{event-W}
W_{k}^{n} := \Bigg\{ \omega: \, \sharp \Big\{i: E_{i}^{\omega,\left[ (k e^{\alpha n^{\gamma}})^{1/d}
\right]_{+}}
< \frac{1}{n}\Big\} = k \Bigg\} \ .
\end{eqnarray}
Let $\omega \in (V_{k}^{n} \setminus W_{k}^{n})$. Then by
$h_{\left[ ((k+1) e^{\alpha n^{\gamma}})^{1/d} \right]_{+}}^{\omega} \leqslant
h_{\left[ (k e^{\alpha n^{\gamma}} \right)^{1/d}]_{+}}^{\omega}$ we get:
\begin{eqnarray*}
\sharp \Big\{i: E_{i}^{\omega,\left[ ((k+1)e^{\alpha n^{\gamma}})^{1/d} \right]_{+}} <
\frac{1}{n}\Big\} \,\geqslant \,  \sharp \Big\{i: E_{i}^{\omega,\left[ (k e^{\alpha n^{\gamma}})^{1/d}
\right]_{+}} <
\frac{1}{n}\Big\} \geqslant k+1  \ .
\end{eqnarray*}
Hence, $(V_{k}^{n} \setminus W_{k}^{n}) \subset V_{k+1}^{n}$, and therefore we have for any fixed $n$ and $k$:
\begin{eqnarray}\label{Recursive-inclusion-V}
V_{k}^{n} \, \subset \, W_{k}^{n} \cup V_{k+1}^{n} \ .
\end{eqnarray}
Applying this inclusion $M$ times, for $k = 1, \dots, M$, we obtain:
\begin{eqnarray}\label{Inclusion}
\bigcup_{k=1}^{M} V_{k}^{n} \subset \Big(W_{1}^{n} \cup
\bigcup_{k=2}^{M} V_{k}^{n}\Big) \subset \Big(W_{1}^{n} \cup W_{2}^{n} \cup
\bigcup_{k=2}^{M} V_{k}^{n}\Big)\subset \dots \subset \left(\bigcup_{k=1}^{M} W_{k}^{n}\right)
\cup V_{M+1}^{n} \ .
\end{eqnarray}
Then we take the limit $M \rightarrow \infty$ to recover the infinite union that one needs in (\ref{Proba-V})
and we use the inclusion (\ref{Inclusion}) to find the inequality:
\begin{eqnarray}\label{Proba-V-2}
\mathbb{P} \big(\bigcup_{k\geq 1} V_{k}^{n} \big) &=&
\lim_{M \rightarrow \infty} \mathbb{P} \big(\bigcup_{k=1}^{M} V_{k}^{n} \big)\\
&\leqslant&  \lim_{M \rightarrow \infty} \Big( \sum_{k=1}^{M}  \mathbb{P} \big( W_{k}^{n} \big) +
\mathbb{P} (V_{(M+1)}^{n})  \Big)  =  \sum_{k=1}^{\infty}  \mathbb{P} \big( W_{k}^{n} \big) +
\lim_{M \rightarrow \infty} \mathbb{P} (V_{M}^{n}).\nonumber
\end{eqnarray}
The limit in the last term can be calculated directly :
\begin{eqnarray}\label{LimLastTerm}
&\,&\lim_{M \rightarrow \infty} \mathbb{P} (V_{M}^{n})
= \lim_{M \rightarrow \infty} \mathbb{P} \Bigg\{ \omega: \,
\sharp \Big\{i: E_{i}^{\omega,\left[ (M e^{\alpha n^{\gamma}})^{1/d} \right]_{+}}
< \frac{1}{n}\Big\} \geqslant M \Bigg\}  \\
&=& \lim_{M \rightarrow \infty} \mathbb{P} \Bigg\{ \omega: \,\nu_{\left[
(M e^{\alpha n^{\gamma}})^{1/d} \right]_{+}}^{\omega} ({1}/{n}) \geqslant \frac{M}
{\left[ (M e^{\alpha n^{\gamma}})^{1/d} \right]_{+} ^{d}} \Bigg\}
= \mathbb{P} \Bigg\{ \omega: \,\nu ({1}/{n})  \geqslant K e^{-\alpha n^{\gamma}}\Bigg\}  \nonumber \ ,
\end{eqnarray}
for some constant $K$. In the last step we used dominated convergence theorem.

Now we can use the Lifshitz tails representation for the asymptotics of the  a.s. non-random limiting IDS,
$\nu(E)$, see (\ref{Lifshitz-tails-infinite-volume}), which implies:
\begin{eqnarray}\label{AsymptLimLastTerm}
\limsup_{n \rightarrow \infty} \, e^{a  n^{d/2}} \, \nu ({1}/{n}) \, \leq \, 1 \ ,
\end{eqnarray}
for  $a>0$. Since we assumed that $0 < \gamma < d/2$, there exists $n_{0} < \infty$
such that by (\ref{LimLastTerm}) and (\ref{AsymptLimLastTerm}) for all $n > n_{0}$ we get:
\begin{eqnarray*}
\lim_{M \rightarrow \infty} \mathbb{P} (V_{M}^{n}) = 0.
\end{eqnarray*}
This last result, along with (\ref{Proba-V}) and (\ref{Proba-V-2}), implies that:
\begin{eqnarray}\label{Proba-W}
\sum_{n > n_{0}} \, \mathbb{P} \Big( \bigcup_{l \geq l_0} \, S_{l}^{n} \Big) \, \leqslant \,
\sum_{n > n_{0}} \, \sum_{k=1}^{\infty} \,  \mathbb{P} \big( W_{k}^{n} \big).
\end{eqnarray}

Now, we show that the upper bound in (\ref{Proba-W}) is finite. First we split the box
$\Lambda_{\left[ (k e^{\alpha n^{\gamma}})^{1/d} \right]_{+}}$  into $m(k,n)$ disjoints sub-cubes of
the side $l(k,n)$,
with the following choice of parameters:
\begin{eqnarray*}
m(k,n) \, &:=& \, \left[ k M_{n} \right]_{+} , \quad M_{n} := B^{-d/2} e^{\alpha n^{\gamma}} n^{-d/2}\\
l(k,n) \, &:=& \, \frac{\left[ (k e^{\alpha n^{\gamma}})^{1/d} \right]_{+}}{(m(k,n))^{1/d}} \ .
\end{eqnarray*}
Here $B$ is the constant that comes from  Lemma \ref{upper-bound-eigenvalue1}. Now by the
Dirichlet-Neumann inequality,
see e.g. \cite{RS-IV}, we get:
\begin{eqnarray}\label{D-N-bracketing}
h_{\left[ (k e^{\alpha n^{\gamma}})^{1/d} \right]_{+}}^{D} \, \geqslant
\, h_{\left[ (k e^{\alpha n^{\gamma}})^{1/d} \right]_{+}}^{N}
\, \geqslant \,\bigoplus_{j=1}^{m(k,n)} h_{l(k,n)}^{j,N},
\end{eqnarray}
where $h_{l(k,n)}^{j,N}$ denotes the Schr\"{o}dinger operator defined in the $j$-th sub-cube of the
side $l(k,n)$, with
Neumann boundary conditions. Note that, by the \textit{positivity} of the random potential, we obtain:
\begin{eqnarray}\label{2-nd-eigenvalue}
E_{j, 2}^{\omega, N} \, \geqslant \, \varepsilon_{j, 2}^{N} \, \geqslant \, \frac{\pi}{l(k,n)^{2}} \,
\geqslant \, \frac{1}{n} \ .
\end{eqnarray}
Here $E_{j, 2}^{\omega, N}$ denotes the \textit{second eigenvalue} of the operator $h_{l(k,n)}^{j,N}$, and
$\varepsilon_{j, 2}^{N}$ the \textit{second eigenvalue} of $-\Delta_{l(k,n)}^{j,N} $ , i.e. the
kinetic-energy operator
defined in the $j$-th sub-cube of the side $l(k,n)$ with the Neumann boundary conditions.

{By equation (\ref{2-nd-eigenvalue}), we know that to estimate the probability of the set (\ref{event-W})
by using the Dirichlet-Neumann inequality (\ref{D-N-bracketing}), only the \textit{ground state} of each
operator $h_{l(k,n)}^{j,N}$ is relevant. Since the sub-cubes are \textit{stochastically independent}, we have:}
\begin{eqnarray*}
\mathbb{P} \big( W_{k}^{n} \big) \, \leqslant \, \mathbb{P} \Bigg\{\omega: \sharp \big\{j:
E_{j, 1}^{\omega, N} <
{1}/{n}\big\} = k \Bigg\} \, \leqslant \, {}^{m(k,n)}C_{k} \, q^{k} (1 - q)^{m(k,n)-k} \, \leqslant \,
{}^{m(k,n)}C_{k} \, q^{k}
\end{eqnarray*}
with $q$ being the probability $\mathbb{P}\{\omega:E_{j, 1}^{\omega, N} < {1}/{n}\}$.
The latter can be estimated by Lemma \ref{upper-bound-eigenvalue1}. So, finally we obtain the
upper bound:
\begin{eqnarray}\label{upper-bound-proba-W}
\mathbb{P} \big( W_{k}^{n} \big) \, \leqslant \, {}^{m(k,n)}C_{k} \, \exp\{-k A (l(k,n))^d\} \ .
\end{eqnarray}
Using Stirling's inequalities, see \cite{F1}:
\begin{eqnarray*}
(2\pi)^{1/2} n^{n+1/2} e^{-n} \, \leqslant \, n! \, \leqslant \, 2(2\pi)^{1/2} n^{n+1/2} e^{-n} \ .
\end{eqnarray*}
we can give an upper bound for the binomial coefficients ${}^{m(k,n)}C_{k}$ in the form:
\begin{eqnarray}\label{upper-bound-c^m_k}
\displaystyle{\frac{2(2\pi)^{\frac{1}{2}} (k M_{n} + \delta)^{(k M_{n} + \delta +1/2)} \exp(-k M_{n} +
\delta)}{(2\pi) k^{k+\frac{1}{2}} \exp (-k)\cdot(k M_{n} + \delta - k)^{(k M_{n} +
\delta-k +1/2)}\exp(-k M_{n}+\delta-k) }} \ ,
\end{eqnarray}
where $\delta \geqslant 0$ is defined by:
\begin{eqnarray*}
 m(k,n) \, = \, \left[ k M_{n} \right]_{+} \, = \, k M_{n} + \delta \ .
\end{eqnarray*}
Then (\ref{upper-bound-c^m_k}) implies the estimate:
\begin{eqnarray*}
{}^{m(k,n)}C_{k} \, \leqslant \, K_{1} \frac{(k M_{n} + \delta)^{k M_{n} + \delta +1/2}}{ k^{k+\frac{1}{2}}
(k M_{n} - k)^{k M_{n} + \delta - k +1/2}} \, \leqslant \, K_{1} (M_{n})^{k}
\displaystyle{\Big(\frac{(1+\sigma_{1})^{(k M_{n} + \delta + \frac{1}{2})}}{(1-\sigma_{2})^{(k M_{n} +
\delta + \frac{1}{2} -k)}}\Big)},
\end{eqnarray*}
for some $K_{1}> 0$ and
\begin{eqnarray*}
\sigma_{1} := \delta (k M_{n})^{-1} , \,\, \sigma_{2} := M_{n}^{-1} \ .
\end{eqnarray*}
Since $\delta/k < 1$ and $\sigma_{1,2} \rightarrow 0$ as $n \rightarrow \infty$, and also using the fact
that
$x \ln (1+1/x) \rightarrow 1$ as $x \rightarrow \infty$, we can find a constant $c>0$ such that,
for $n$ large enough one gets the estimate:
\begin{eqnarray}\label{upper-bound-c^m_k-2}
{}^{m(k,n)}C_{k}  \, \leqslant \, K_{1} (M_{n})^{k} \displaystyle{\Big(\frac{(1+M_{n}^{-1})^{(k M_{n})}}
{(1- M_{n}^{-1})^{(k M_{n} - k)}}\Big)} \, \leqslant \, K_{1} (M_{n})^{k} \, e^{ck} \ .
\end{eqnarray}
The side $l(k,n)$ of sub-cubes has a lower bound :
\begin{eqnarray}\label{upper-bound-lkn}
l(k,n) =  \frac{\left[ (k e^{\alpha n^{\gamma}})^{1/d} \right]_{+}}{(m(k,n))^{1/d}} \geqslant
\frac{(k e^{\alpha n^{\gamma}})^{1/d}}{({k} e^{\alpha n^{\gamma}} (B n)^{-d/2} +
\delta)^{1/d}} \geqslant \Big(B^{d/2} \, n^{d/2} \, \frac{1}{1+\sigma_{1}}\Big)^{1/d} \ .
\end{eqnarray}
Combining (\ref{upper-bound-c^m_k-2}), (\ref{upper-bound-lkn}) and (\ref{upper-bound-proba-W}) we obtain
a sufficient upper bound:
\begin{eqnarray*}
\sum_{k\geqslant 1} \mathbb{P} \big( W_{k}^{n} \big) &\leqslant&\sum_{k\geqslant 1} {}^{m(k,n)}C_{k} \,
e^{-k A l^{d}(k,n)}\\
&\leqslant& \sum_{k\geqslant 1} \,  K_{1} \, (M_{n})^{k} \, e^{c k} \,
e^{-k \, A \, B^{d/2} \, n^{d/2}/(1+\sigma_{1})}\\
&\leqslant& K_{2} \sum_{k\geqslant 1} \exp \Big\{ k\Big( \alpha n^{\gamma} - (d/2)\ln (n B)+ c
-AB^{d/2} n^{d/2}   \Big) \Big\}\\
&\leqslant& K_{3} \sum_{k\geqslant 1}  \exp{k\Big( \alpha n^{\gamma} -AB^{d/2} n^{d/2} + K_{4}\Big)} \,
\leqslant \, K_{5} \, \exp (-K_{6} n^{d/2}) \ .
\end{eqnarray*}
Here $K_{i}$ are some finite, positive constants independent of $k$, $n$, $l$, for any $n$ large enough.
Now the lemma immediately follows from (\ref{Proba-W}).  \hfill $\square$

\noindent \textbf{Proof of Theorem \ref{Lifshitz-tails}}:
{Let $A_{n}$ to be the event:
\begin{eqnarray}\label{event A}
A_{n} \, :=  \,  \left\{\omega: \nu_{l}^{\omega} ({1}/{n}) > \,
e^{-\alpha n^{\gamma}}\quad \textrm{for some} \,\, l \right\}.
\end{eqnarray}
By Lemma \ref{lemma-proof-Lifshitz-tails}, we have:
\begin{eqnarray*}
\sum_{n \geqslant 1} \, \mathbb{P} \big(A_{n}\big) \,\, < \,\, \infty,
\end{eqnarray*}
and therefore, by the Borel-Cantelli lemma one gets
that with probability one, only a \textit{finite} number of events $A_{n}$ occur.
In other words, there is a subset $\widetilde{\Omega}\subset\Omega$ of full measure, $\mathbb{P}(\widetilde{\Omega}) =1$,
such that for any $\omega\in \widetilde{\Omega}$ one can find a \textit{finite} and independent on $l$
number $n_0(\omega) < \infty$ for which, in contrast to (\ref{event A}), we have:
\begin{eqnarray*}
\nu^{\omega}_{l} (1/n) \, \leqslant \, e^{-\alpha n^{\gamma}}, \quad \textrm{for all} \ \, n > n_0(\omega)\
\textrm{and for all}\ l\geqslant 1.
\end{eqnarray*}}
Define $\mathcal{E}(\omega) := 1/n_0(\omega)$. For any $E \leqslant \mathcal{E}(\omega)$, we can find
$n \geqslant n_0(\omega)$ such that:
\begin{eqnarray*}
\frac{1}{2n} \, \leqslant \, E \, \leqslant \, \frac{1}{n},
\end{eqnarray*}
and the theorem follows with the constant $\alpha$  modified by a factor $2^{-\gamma}$.
\hfill $\square$

\section{On the nature of the generalized condensates in the Luttinger-Sy model}\label{Sect-g-BEC-Lutt-Sy}
\setcounter{equation}{0}
\renewcommand{\theequation}{\arabic{section}.\arabic{equation}}
In this section, we study the van den Berg-Lewis-Pul\'{e} classification of generalized BE
condensation (see discussion in Section \ref{section-gBEC-random-eigenstates}) in a particular case
of the so-called Luttinger-Sy model with point impurities \cite{LS}.

Let $u(x)\geq 0,\,x\in \mathbb{R}$, be a continuous function with a \textit{compact} support called a
(\textit{repulsive}) single-impurity potential. Let
$\left\{ \mu_{\lambda }^{\omega }\right\} _{\omega \in \Omega }$ be the \textit{random} Poisson
measure on $\mathbb{R}$ with intensity $\lambda >0$ :
\begin{equation}\label{Ran-Poiss-mes}
\mathbb{P}\left( \left\{ \omega \in \Omega :\mu _{\lambda
}^{\omega }(\Lambda )=n\right\} \right) =\frac{\left( \lambda
\left\vert \Lambda \right\vert\right)^{n}}{n!}e^{-\lambda \left\vert \Lambda \right\vert }\,
\ , \ \ n\in \mathbb{N}_{0}=\mathbb{N}\cup \left\{ 0\right\} ,
\end{equation}
for any bounded Borel set $\Lambda \subset \mathbb{R}$. Then the
non-negative random potential $v^{\omega}$ generated by the Poisson distributed
local impurities has realizations
\begin{equation}
v^{\omega }(x):=\int_{\mathbb{R}}\mu_{\lambda }^{\omega}(dy)u(x-y)=\sum_{x_{j}^\omega\in X^\omega}
u(x-x_{j}^{\omega }) \ .
\label{Ran-Poiss-pot}
\end{equation}%
Here the random set $X^\omega$ corresponds to impurity positions
$X^\omega = \left\{ x_{j}^{\omega }\right\}_j \subset \mathbb{R}$,
which are the atoms of the random point Poisson measure, i.e.,
$\sharp \, \{X^{\omega}\upharpoonright \Lambda\}= \mu_{\lambda}^{\omega}(\Lambda)$ is the number of
impurities in the set $\Lambda$. Since the expectation
$\mathbb{E}\left(\nu_{\lambda}^{\omega}(\Lambda) \right)=
\lambda\left|\Lambda\right|$, the parameter $\lambda$ coincides with the \textit{density} of impurities
on $\mathbb{R} $.

Luttinger and Sy defined their model by restriction of the single-impurity potential to the
case of point $\delta$-potential with amplitude $a \rightarrow + \infty$. Then the corresponding random
potential (\ref{Ran-Poiss-pot}) takes the form:
\begin{equation}\label{ran-delta1}
v_{a}^{\omega }(x):=\int_{\mathbb{R}}\nu _{\lambda}^{\omega
}(dy) a \delta (x-y) = a \, \sum_{x_{j}^\omega\in X^\omega}\delta
(x-x_{j}^{\omega }) \ .
\end{equation}
Now the self-adjoint one-particle random Schr\"{o}dinger operator
$h_{a}^{\omega}:= h^{0}\dotplus v_{a}^{\omega}$
is defined in the sense of the sum of quadratic forms (\ref{Schrodinger-operator-inf}). The strong resolvent
limit
$h_{LS}^{\omega}:= s.r.\lim_{a \rightarrow +\infty} h_{a}^{\omega}$  is the Luttinger-Sy model.

Since $X^{\omega}$ generates a set of intervals
$\left\{I_{j}^{\omega}:=(x_{j-1}^{\omega},x_{j}^{\omega})\right\}_j$
of lengths $\left\{L_{j}^{\omega}:= x_{j}^{\omega} - x_{j-1}^{\omega}\right\}_j$, one gets decompositions
of the one-particle Luttinger-Sy Hamiltonian:
\begin{equation}\label{ham-orthg-sum}
h_{LS}^{\omega} =
\bigoplus_{j}h_{D}(I_{j}^{\omega}) \ , \  \  \ {\rm{dom}}(h_{LS}^{\omega}) \subset
\bigoplus_{j}L^2(I_{j}^{\omega}) \
\ , \ \ \omega\in\Omega  \ ,
\end{equation}
into random disjoint free Schr\"{o}dinger operators $\left\{h_D(I_{j}^{\omega})\right\}_{j , \omega}$
with \textit{Dirichlet} boundary conditions at the end-points of intervals
$\left\{I_{j}^{\omega}\right\}_j \ $.
Then the  Dirichlet restriction $h_{l,D}^{\omega}$ of the Hamiltonian $h_{LS}^{\omega}$ to a fixed  interval
$\Lambda_{l}= (-l/2, l/2)$ and the corresponding change of notations are evident: e.g.,
$\left\{I_{j}^{\omega}\right\}_j  \mapsto \left\{I_{j}^{\omega}\right\}_{j=1}^{M^{l}(\omega)}$, where
$M^{l}(\omega)$ is total number of subintervals in  $\Lambda_{l}$ corresponding to the set $X^{\omega}$.
For rigorous definitions and some results concerning this model we refer the reader to \cite{LZ}.

Since this particular choice of random potential is able to produce Lifhsitz tails in the sense of
(\ref{Lifshitz-tails-infinite-volume}), see Proposition 3.2 in \cite{LZ}, it follows that such a model exhibits a
generalized BEC in random eigenstates, see (\ref{usual-BEC}). In fact, it was shown in \cite{LZ} that \textit{only} the
random ground state $\phi_{1}^{\omega,l}$ of $h_{l,D}^{\omega}$ is \textit{macroscopically} occupied. In our notations this
means that
\begin{eqnarray}\label{BEC-typeI-random-eigenstates-Luttinger-Sy-model}
\lim_{l \rightarrow \infty} \, \frac{1}{l} \, \langle N_{l}(\phi_{1}^{\omega,l})\rangle_{l} \, &=& \,
\left\{ \begin{array}{ll}
              0 & \,\,\textrm{if} \,\, \overline{\rho} <  \rho_{c}\\
               \overline{\rho} -\rho_{c} \,\,&\,\,  \textrm{if} \,\,  \overline{\rho} \geqslant  \rho_{c}
              \end{array} \right.\\
\lim_{l \rightarrow \infty} \, \frac{1}{l} \, \langle N_{l}(\phi_{i}^{\omega,l})\rangle_{l} \, &=& \, 0,
\quad \textrm{for all}\,\, i > 1  \ . \nonumber
\end{eqnarray}
According to the van den Berg-Lewis-Pul\'{e} classification this corresponds to the \emph{type I}
Bose-condensation in the random eigenstates $\{\phi_{i}^{\omega}\}_{i\geq 1}$.

Following the line of reasoning of Section \ref{Sect-kinetic-occupation-measure}, we now consider the
corresponding BEC in the kinetic-energy eigenstates. We retain the notation used in that section and explain briefly
the minor changes required in the application of our method to the Luttinger-Sy model.

We first state the equivalent of Theorem \ref{existence-occupation-measure-and-BEC} for this particular model.
\begin{theorem}\label{kinetic-occupation-measure-Luttinger-Sy}
Theorem \ref{existence-occupation-measure-and-BEC} holds with the function $g$ defined as follows
\begin{eqnarray*}
g(k) &=& \frac{1}{(2\pi)^{d/2}} \, \int_{\mathbb{R}^{d}} \, \ud x \, e^{ikx} \, \sum_{n \geqslant 1}
e^{n\beta\mu_{\infty}} \frac{e^{-\|x\|^2(1/2n\beta)}}{(2\pi n\beta)^{d/2}} \times\\
&\times&  \int_{\Omega_{(0,x)}^{n\beta}} \, w^{n\beta}(\ud\xi) \,
\exp\Big(-\lambda  \big(\sup_{s} \xi(s) - \inf_{s} \xi(s)  \big) \Big).
\end{eqnarray*}
\end{theorem}
The scheme of the proof is the same as above, cf. Sections \ref{Sect-kinetic-occupation-measure} and
\ref{Sect-Lifshitz-tails}. First, we note that Lemmas \ref{lower-bound-kineticBEC}
and \ref{absolute-continuity-kinetic-occupation-measure} apply immediately. The positivity of the random potential
has to be understood in terms of quadratic forms, see (\ref{Positive-Potential}).

Before continuing, we need to highlight a minor change concerning the \textit{finite-volume} Lifshitz tails arguments.
Although} the Theorem \ref{Lifshitz-tails} is valid for the Luttinger-Sy model, its proof 
(see Section \ref{Sect-Lifshitz-tails})
requires a minor modification, as the assumption of Lemma \ref{upper-bound-eigenvalue1} is clearly
not satisfied for the case of singular potentials. However, by direct calculation we can obtain the same estimate
with the constant $B=\pi^2/4$ in (\ref{estim-lemma5.1}).
First, suppose that there is at least one impurity in the box, then the eigenvalues will be of the form (for some $j$)
\begin{eqnarray*}\label{Luttinger-Sy-inner-interval-eigenvalue}
(n^2 \pi^2)/{(L_{j}^{\omega})^2}, \quad n=1,2,\dots
\end{eqnarray*}
if $I_{j}^{\omega}$ is an inner interval (that is, its two endpoints correspond to impurities), and
\begin{eqnarray*}\label{Luttinger-Sy-outer-interval-eigenvalue}
((n+1/2)^2 \pi^2)/(L_{j}^{\omega})^2, \quad n=0,1,2,\dots .
\end{eqnarray*}
if $I_{j}^{\omega}$ is an outer interval (that is, one endpoint corresponds to an impurity, and the other one to
the boundary of $\Lambda_{l}$). Therefore, $E_1^{\omega,l,N} \geqslant B/l^2$ since obviously $L_{j}^{\omega} < l$.
Now, if there is no impurity in the box $\Lambda_{l}$, then $E_{1}^{\omega,\, l, N} = 0<B/l^2$. But due to the Poisson
distribution (\ref{Ran-Poiss-mes}) this happens with probability $e^{-\lambda l}$, proving the same estimate as in
Lemma \ref{upper-bound-eigenvalue1}.\\
With this last observation, the proof of the Theorem \ref{Lifshitz-tails} in Section \ref{Sect-Lifshitz-tails}
can be carried out verbatim, without any further changes.

Our next step is to split the measure $\tilde{m}_{l}$ into two, $\tilde{m}^{(1)}_{l}$ and $\tilde{m}^{(2)}_{l}$, see
(\ref{occupation-measure-with-cutoff}), and prove the statement equivalent to the Theorem
\ref{Laplace-transform-occupation-measure}.
\begin{theorem}\label{Laplace-transform-occupation-measure-Luttinger-Sy}
For any $d \geqslant 1$, the sequence of Laplace transforms of the measures $\tilde{m}_{l}^{(1)}$:
\begin{equation*}
f_{l}(t; \beta, \mu_{l}) \,\,\, := \,\,\,  \int_{\mathbb{R}} \, \tilde{m}_{l}^{(1)}(\ud \varepsilon) \ e^{-t\varepsilon}
\end{equation*}
converges for any $t > 0$ to a (non-random) limit $f(t; \beta, \mu_{\infty})$ , which is given by:
\begin{eqnarray*}
f(t; \beta, \mu_{\infty}) \, &=& \, \sum_{n \geqslant 1} e^{n\beta\mu_{\infty}} \, \int_{\mathbb{R}^{d}}
\ud x \, \frac{e^{-\|x\|^2(1/2n\beta+1/2t)}}{(4\pi^2tn\beta)^{d/2}}\\
&\times& \int_{\Omega_{(0,x)}^{n\beta}} \, w^{n\beta} (\ud \xi) \,
\exp \Big(-\lambda \big(\sup_{s} \xi(s) - \inf_{s} \xi(s)  \big)\Big).
\end{eqnarray*}
\end{theorem}
\textbf{Proof:}
We follow the proof of Theorem \ref{Laplace-transform-occupation-measure}, using the same notation.
The uniform convergence is obtained the same way, since the bounds (\ref{upper-bound-a_l(n)-rho-less-rho_c}),
(\ref{upper-bound-a^1_n}), and (\ref{upper-bound-a^2_n}) are also valid in this case.
As in (\ref{lim a}), we can use the ergodic theorem to obtain:

\begin{eqnarray}\label{lim a-inf}
\lim_{l \rightarrow \infty} a_{l} (n) = e^{n\beta\mu_{\infty}} \mathbb{E}_{\omega} \,  \int_{\mathbb{R}}\ud x \,
\frac{e^{-\|x\|^2(1/2n\beta+1/2t)}}{(4\pi^2tn\beta)^{d/2}}\sum_{j} \int_{\Omega_{(0,x)}^{n\beta}} \,
w^{n\beta}(\ud \xi)  \, \chi_{{I_{j}}^{\omega}, n\beta } (\xi).
\end{eqnarray}
We have used the fact that the Dirichlet boundary conditions at the impurities split up the space $\mathcal{H}_l$
into a direct sum
of Hilbert spaces (see (\ref{ham-orthg-sum})). This can be seen from the expression
\begin{eqnarray*}
\lim_{l \rightarrow \infty} a_{l} (n) = e^{n\beta\mu_{\infty}} \int_{\mathbb{R}}\ud x \,
\frac{e^{-\|x\|^2(1/2n\beta+1/2t)}}{(4\pi^2tn\beta)^{d/2}} \ \mathbb{E}_{\omega} \int_{\Omega_{(0,x)}^{n\beta}}
w^{n\beta}(\ud \xi)  \, \, e^{-\int_{0}^{n\beta} \, \ud s \, a \sum_{x_{j}^\omega\in X^\omega}\delta
(\xi(s)-x_{j}^{\omega })}.
\end{eqnarray*}
by formally putting the amplitude, $a$, of the point impurities (\ref{ran-delta1}) equal to $+\infty$.
Because of the characteristic functions $\chi_{I_{j}^{\omega},n\beta}$, which constrain the paths $\xi$ to remain in the
interval $I_{j}^{\omega}$ in time $n\beta$, the sum in (\ref{lim a-inf})
reduces to only \textit{one} term:
\begin{eqnarray}\label{expectation-Laplace-transform-Luttinger-Sy}
\lim_{l \rightarrow \infty} a_{l} (n)  = e^{n\beta\mu_{\infty}}  \int_{\mathbb{R}}\ud x \,
\frac{e^{-\|x\|^2(1/2n\beta+1/2t)}}{(4\pi^2tn\beta)^{d/2}} \  \mathbb{E}_{\omega}
\int_{\Omega_{(0,x)}^{n\beta}} \, w^{n\beta}(\ud \xi) \, \chi_{(a_{\omega}, b_{\omega}),n\beta }(\xi) \ ,
\end{eqnarray}
where $(a_{\omega}, b_{\omega})$, is the interval among the $I_{j}^{\omega}$'s which contains $0$.

The expression in (\ref{expectation-Laplace-transform-Luttinger-Sy}) can be simplified further by computing the
expectation $\mathbb{E}_{\omega}$ explicitly.

\noindent First, note that the Poisson impurity positions: $a_{\omega}, b_{\omega}$ are independent random variables
{and} by definition,
$a_\omega$ is negative while $b_\omega$ is positive.
For the random variable $b_{\omega}$ the distribution function is:
\begin{eqnarray*}
\mathbb{P}\ (b_{\omega} < b):= \mathbb{P}\{(0,b)\ \textrm {contains at least one impurity} \}= 1 - e^{-\lambda b},
\end{eqnarray*}
and therefore its probability density is $\lambda e^{-\lambda b}$ on $(0,\infty)$. Similarly for $a_{\omega}$ one gets:
\begin{eqnarray*}
\mathbb{P}\ (a_{\omega} < a):= \mathbb{P}\{(a,0)\ \textrm {contains no impurities} \}= e^{-\lambda |a|} \, =
\, e^{\lambda a},
\end{eqnarray*}
and thus its density is $ \lambda e^{\lambda a}$ on $(-\infty, 0)$. Using these distributions in
(\ref{expectation-Laplace-transform-Luttinger-Sy}) we obtain:
\begin{eqnarray*}
\lim_{l \rightarrow \infty} a_{l} (n)
&=& e^{n\beta\mu_{\infty}} \lambda^{2}  \int_{-\infty}^{0} \ud a \, e^{\lambda a} \int_{0}^{\infty} \ud b \, e^{-\lambda b}
\int_{\mathbb{R}}\ud x \, \frac{e^{-\|x\|^2(1/2n\beta+1/2t)}}{(4\pi^2tn\beta)^{d/2}}\times\\
&\,&\times \int_{\Omega_{(0,x)}^{n\beta}} \, w^{n\beta}(\ud \xi)  \, \chi_{(a, b)}(\xi) \\
&=& e^{n\beta\mu_{\infty}} \lambda^{2}  \int_{-\infty}^{0} \ud a \, e^{\lambda a} \int_{0}^{\infty} \ud b \, e^{-\lambda b}
\int_{\mathbb{R}}\ud x \, \frac{e^{-\|x\|^2(1/2n\beta+1/2t)}}{(4\pi^2tn\beta)^{d/2}}\times\\
&\,&\times  \int_{\Omega_{(0,x)}^{n\beta}} \, w^{n\beta}(\ud \xi)  \,  \mathbf{1}(\sup_{s} (\xi(s)) \leqslant b)  \,
\mathbf{1}(\inf_{s} (\xi(s)) \geqslant a)\\
&=& e^{n\beta\mu_{\infty}} \lambda^{2}\int_{\mathbb{R}}\ud x \, \frac{e^{-\|x\|^2(1/2n\beta+1/2t)}}
{(4\pi^2tn\beta)^{d/2}}\times\\
&\,&\times \int_{\Omega_{(0,x)}^{n\beta}} \, w^{n\beta}(\ud \xi) \, \int_{-\infty}^{\inf_{s} (\xi(s))}
\ud a \, e^{\lambda a}
\int_{\sup_{s} (\xi(s))}^{\infty} \ud b \, e^{-\lambda b} \ ,\\
\end{eqnarray*}
and the Theorem \ref{Laplace-transform-occupation-measure-Luttinger-Sy} follows by explicit computation of the last
two integrals. \hfill $\square$

\noindent{\textbf{Proof of Theorem \ref{kinetic-occupation-measure-Luttinger-Sy}:}
Having proved Theorem \ref{Laplace-transform-occupation-measure-Luttinger-Sy},
it is now straightforward to derive the analogue of Corollary \ref{existence-momentum-occupation-measure} for the
Luttinger-Sy model. Note also that the Corollary \ref{weight-measure-m1} remains unchanged, since only the uniform
convergence was used. With these results, the proof of Theorem \ref{kinetic-occupation-measure-Luttinger-Sy}
follows in the same way as for Theorem \ref{existence-occupation-measure-and-BEC}.  \hfill $\square$}

We have proved, in Theorem \ref{kinetic-occupation-measure-Luttinger-Sy}, that the Luttinger-Sy model exhibits
g-BEC in the kinetic energy states. But, in this particular case, we can go further and determine the particular
\emph{type} of g-BEC in the kinetic energy states. Recall that the g-BEC
in the \emph{random} eigenstates is only in the \textit{ground} state, that is, of the \textit{type} I, see
(\ref{BEC-typeI-random-eigenstates-Luttinger-Sy-model}) and \cite{LZ} for a comprehensive review.
Here we shall show that the g-BEC in the kinetic-energy eigenstates is in fact of the \textit{type} III, namely:
\begin{theorem}\label{classification-kineticBEC-Luttinger-Sy-model}
In the Luttinger-Sy model none of the kinetic-energy eigenstates is macroscopically occupied:
\begin{eqnarray*}
\lim_{l \rightarrow \infty} \,\frac{1}{l} \, \langle N_{l}(\psi_{k})\rangle_{l} \, = \, 0
\quad \textrm{\emph{for all}} \,\, k\in \Lambda_{l}^*,
\end{eqnarray*}
even though for $\overline{\rho} >\rho_c$ there is a generalized BEC.
\end{theorem}
To prove this theorem we shall exploit the finite-volume localization properties of the
random eigenfunctions $\phi_{i}^{\omega,l}$ of the Hamiltonian $h_{l,D}^{\omega}$.
Since the impurities split up the box $\Lambda_{l}$ into a finite
number $M^{l}(\omega)$ of sub-intervals $\left\{I_{j}^{\omega}\right\}_{j=1}^{M^{l}(\omega)}$,
{by virtue of the corresponding orthogonal decomposition of $h_{l,D}^{\omega}$, cf (\ref{ham-orthg-sum}),
the normalized random eigenfunctions $\phi_{s}^{\omega,l}$ are in fact \textit{sine-waves} with supports
in each of these sub-intervals and thus satisfy:
\begin{eqnarray}\label{explicit-form-random-eigenfunction-Luttinger-Sy-model}
{|\phi_{s}^{\omega,l}(x)| \, < \,  \ \sqrt{\frac{2}{L_{j_{s}}^{\omega}}}\,\mathbf{1}_{I_{j_{s}}^{\omega}}(x) \ \  ,
 \ \  1\leq j_{s}\leq M^{l}(\omega)} \ .
\end{eqnarray}
We require an estimate of the size $L_{j}^{\omega}$ of these random sub-intervals, which we
obtain in the following lemma.
\begin{lemma}\label{lemma-localization-Luttinger-Sy-model}
Let $\lambda > 0$ be a mean concentration of the point Poisson impurities on $\mathbb{R}$.
Then eigenfunctions $\phi_{j}^{\omega}$ are localized in sub-intervals of logarithmic size, in
the sense that
for any $\kappa > 4$, one has a.s. the estimate:
\begin{eqnarray*}
\limsup_{l \rightarrow \infty} \frac{\max_{1\leqslant j \leqslant M^{l}(\omega)} \,
L_{j}^{\omega}}{\ln l} \, \leqslant \, \frac{\kappa}{\lambda}.
\end{eqnarray*}
\end{lemma}
\textbf{Proof}: Define the set
\begin{eqnarray*}
S_{l} := \big\{ \omega:  \max_{1\leqslant j \leqslant M^{l}(\omega)} \, L_{j}^{\omega} \, > \,
\frac{\kappa}{\lambda} \, \ln l  \big\}.
\end{eqnarray*}
Let $n := \big[{2 \lambda l}/{(\kappa \ln l)}\big]_{+}$, and define a new box:
\begin{eqnarray*}
\widetilde{\Lambda}_{l} := [-\frac{n}{2}(\frac{\kappa}{2\lambda} \ln l)\, , \,
\frac{n}{2}(\frac{\kappa}{2\lambda} \ln l)] \, \supset \, \Lambda_{l} \ .
\end{eqnarray*}
Split this bigger box into $n$ identical disjoints intervals
$\{I_{m}^{l}\}_{m = 1}^{n}$ of size ${\kappa}{(2\lambda)^{-1}} \ln l$. If $\omega \in S_{l}$, then there
exists at least one empty interval $I_{m}^{l}$ (interval without any impurities), and therefore
the set
\begin{eqnarray*}
S_{l} \, \subset \, \bigcup_{1 \leqslant m \leqslant n} \{\omega: I_{m}^{l} \ \textrm{is empty} \} \ .
\end{eqnarray*}
By the Poisson distribution (\ref{Ran-Poiss-mes}), the probability for the interval $I_{m}^{l}$ to be
empty depends only on its size, and thus
\begin{eqnarray*}
\mathbb{P}(S_{l}) \, \leqslant \, n \, \exp (-\lambda \frac{\kappa}{2\lambda} \ln l) \, \leqslant \,
\left[\frac{2 \lambda l}{\kappa \ln l}\right]_{+} \ l^{-\kappa/2}.
\end{eqnarray*}
Since we choose $\kappa > 4$, it follows that
\begin{eqnarray*}
\sum_{l \geqslant 1} \mathbb{P}(S_{l}) \, < \, \infty.
\end{eqnarray*}
Therefore, by the Borel-Cantelli lemma, there exists a subset $\widetilde{\Omega}\subset\Omega$
of full measure, $\mathbb{P}(\widetilde{\Omega}) =1$, such that for each $\omega\in \widetilde{\Omega}$
one can find $l_{0}(\omega) < \infty$ with
\begin{eqnarray*}
\mathbb{P} \ \{\omega: \max_{1\leqslant j \leqslant M^{l}(\omega)} \, L_{j}^{\omega} \, \leqslant \,
\frac{\kappa}{\lambda} \, \ln l\} = 1 \ .
\end{eqnarray*}
for all $l \geqslant l_{0}(\omega)$.      \hfill $\square$

Now we can prove the main statement of this section.\\
\noindent \textbf{Proof of Theorem \ref{classification-kineticBEC-Luttinger-Sy-model}:}
The atom of the measure $\tilde{m}$ has already been established in Theorem
\ref{kinetic-occupation-measure-Luttinger-Sy}. Concerning the macroscopic occupation of a single state,
we have
\begin{eqnarray*}
\frac{1}{l} \, \langle N_{l}(\psi_{k})\rangle_{l} &=& \frac{1}{l} \, \sum_{i} |(\phi_{i}^{\omega,l},
\psi_{k})|^{2}
\langle N_{l}(\phi_{i}^{\omega,l})\rangle_{l}\\
&=&  \frac{1}{l} \, \sum_{i}  \langle N_{l}(\phi_{i}^{\omega,l})\rangle_{l} \left|\int_{\Lambda_{l}}
\ud x \ \overline{\psi}_{k} (x) \, \phi_{i}^{\omega,l}(x)\right|^{2}\\
&\leqslant&  \frac{1}{l} \, \sum_{i}  \langle N_{l}(\phi_{i}^{\omega,l})\rangle_{l}  \ \frac{1}{l}
\left(\int_{\Lambda_{l}} \ud x  \, |\phi_{i}^{\omega,l}(x)|\right)^{2},
\end{eqnarray*}
where in the last step we have used the bound $|\psi_{k}| \leqslant 1/\sqrt{l}$ . Therefore, by
(\ref{explicit-form-random-eigenfunction-Luttinger-Sy-model}) and Lemma
\ref{lemma-localization-Luttinger-Sy-model},
we obtain a.s. the following estimate:
\begin{eqnarray*}
\frac{1}{l} \, \langle N_{l}(\psi_{k})\rangle_{l}  \, \leqslant \,  \frac{1}{l} \, \sum_{i}
\langle N_{l}(\phi_{i}^{\omega,l})\rangle_{l} \  \frac{1}{l} \ \frac{\kappa}{\lambda} \, \ln l \ ,
\end{eqnarray*}
which is valid for for large enough $l$  and for any $\kappa > 4$. The theorem then follows by taking the thermodynamic
limit.   \hfill $\square$

\section{Application to weak (scaled) non-random potentials}\label{Sect-appl-weak-potentials}
\setcounter{equation}{0}
\renewcommand{\theequation}{\arabic{section}.\arabic{equation}}
{It is known for a long time, see e.g. \cite{P}, \cite{VdBL}, that BEC can be enhanced in
low-dimensional systems by imposing a weak (scaled) external potential. Recently this was a subject of a new
approach based on the Random Boson Point Field method \cite{TZ}. In this section, we show that, with some
minor modifications our method can be extended to cover also the case of these scaled \textit{non-random} potentials.}

{Let $v$ be a non-negative, continuous real-valued function defined on the closed unit cube
${\overline \Lambda}_1\subset \mathbb{R}^d$.} The \textit{one-particle} Schr\"{o}dinger operator with
a \textit{weak}
(\textit{scaled}) external potential in a box $\Lambda_{l}$ is define by:
\begin{eqnarray}\label{Schrodinger-operator-weak-external-potential}
h_{l} \, = \, -\shalf \Delta_D \, + \, v({x_{1}}/{l},\dots,{x_{d}}/{l}) \ .
\end{eqnarray}
{Let $\{\varphi_{i}^{l},\, E_{i}^{l}\}_{i \geqslant 1}$ be the set of orthonormal eigenvectors and
corresponding
eigenvalues of the operator (\ref{Schrodinger-operator-weak-external-potential}). As usually we put
$E_{1} \leqslant E_{2} \leqslant \dots$ by convention. The many-body Hamiltonian for the perfect Bose gas is
defined in the same way as in Section \ref{model-notations}. We keep the notations $m$ and $\tilde{m}$ for the
occupation measures of the eigenstates $\{\varphi_{i}^{l}\}_{i \geqslant 1}$ and of the kinetic-energy states
respectively.
We denote the \textit{integrated density of states} (IDS) of the Schr\"{o}dinger operator
(\ref{Schrodinger-operator-weak-external-potential}) by $\nu_{l}$, and by
$\nu =\lim_{l\rightarrow\infty}\nu_{l}$
its weak limit. We assume that the first eigenvalue $E_1^{l} \rightarrow 0$
as $l \rightarrow \infty$, which is the case, when e.g. $v(0)= 0$. This assumption is equivalent to condition (iii),
Section \ref{model-notations}. It ensures that for
a given
mean particle density $\overline{\rho}$ the chemical potential $\mu_{\infty}(\beta,\overline{\rho})$
satisfies the
relation (\ref{mu-inf}), where $\overline{\mu}:= \overline{\mu}(\beta,\overline{\rho})$ is a (unique)
solution of
the equation \cite{P}:}
\begin{eqnarray}\label{eq-rho-mu-v}
\overline{\rho} \, = \, \, \sum_{n \geqslant 1} \,  \frac{1}{(2 \pi n \beta)^{d/2}} \, \int_{\Lambda_{1}}
\ud x \, e^{n \beta (\mu - v(x))} \, = \, \int_{[0,\infty)} \nu^{0}(\ud E) \, \int_{\Lambda_{1}} \ud x \,
\left( e^{\beta (E + v(x)- \mu)} - 1 \right)^{-1} \ ,
\end{eqnarray}
for $\overline{\rho} \leq  \rho_{c}$, where the boson critical density is given by:
\begin{eqnarray}\label{critical-density-weak-external-potential}
\rho_{c} \, = \, \sum_{n \geqslant 1} \,  \frac{1}{(2 \pi n \beta)^{d/2}} \, \int_{\Lambda_{1}}
\ud x \, e^{-n \beta v(x)} \, = \, \int_{[0,\infty)} \nu^{0}(\ud E) \, \int_{\Lambda_{1}} \ud x \,
\left( e^{\beta (E + v(x))} - 1 \right)^{-1} \ .
\end{eqnarray}
Here $\nu^{0}$ is the IDS (\ref{density-states-kinetic}) of the kinetic-energy operator
(\ref{Kinetic-energy-operator}).
In particular the value  $\rho_{c} = \infty$ is allowed in
(\ref{critical-density-weak-external-potential}).
If $\rho_{c} < \infty$, the existence of a generalized BEC in the states
$\{\varphi_{i}^{l}\}_{i\geq 1}$ follows by
the same arguments as in Section 3. For example, the choice: $v(x) = |x|$,  makes the critical density
finite even in dimension one, see e.g. \cite{P}.

{Now, we prove the statements equivalent to the Theorem \ref{existence-occupation-measure-and-BEC}:}
\begin{theorem}\label{existence-occupation-measure-and-BEC-weak-potential}
The sequence $\{\tilde{m}_{l}\}_{l\geqslant 1}$ of the one-particle kinetic states occupation measures has a weak limit
$\tilde{m}$ given by:
\begin{eqnarray*}
\tilde{m}(\ud \varepsilon) \, = \,  \left\{ \begin{array}{ll}
               (\overline{\rho} - \rho_{c}) \delta_{0}(\ud \varepsilon)  \, +
               \, F(\varepsilon) \nu^{0}(\ud \varepsilon) & \,, \ \textrm{if}
               \,\, \overline{\rho} \geqslant  \rho_{c} \ ,\\
              F(\varepsilon) \nu^{0}(\ud \varepsilon) \,\,&\,, \  \textrm{if} \,\,  \overline{\rho} <  \rho_{c} \ ,
              \end{array} \right.
\end{eqnarray*}
where the density $F(\varepsilon)$ is defined by:
\begin{eqnarray*}
F(\varepsilon) \, = \,  \int_{\Lambda_{1}} \ud x \ \big(e^{\beta(\varepsilon + v(x) - \mu_{\infty})} - 1\big)^{-1} \ ,
\end{eqnarray*}
and $\mu_{\infty}:=\mu_{\infty}(\beta,\overline{\rho})$ satisfies the relation (\ref{mu-inf}).
\end{theorem}
{We note the similarity of this result with the free Bose gas. Indeed, the kinetic-energy states occupation measure
density is reduced to the free gas one, with the energy shifted by the external potential $v$ and then \textit{averaged }
over the unit cube.}

The proof requires the same tools as in the random case. As before, we split the occupation measure into
two parts:
\begin{eqnarray*}
\tilde{m}_{l} \, &=& \, \tilde{m}_{l}^{(1)} \, + \, \tilde{m}_{l}^{(2)}\qquad \textrm{with}\\
\tilde{m}_{l}^{(1)} (A)\, &:=& \,  \sum_{n \geqslant 1} \,  \frac{1}{V_{l}} \, \textrm{Tr} \, P_{A} \,
(e^{-n\beta(h_{l} - \mu_{l})}) \, \mathbf{1} (\mu_{l} \leqslant 1/n) \ , \nonumber\\
\tilde{m}_{l}^{(2)} (A)\, &:=& \,  \sum_{n \geqslant 1} \,  \frac{1}{V_{l}} \, \textrm{Tr} \, P_{A} \,
(e^{-n\beta(h_{l} - \mu_{l})}) \, \mathbf{1} (\mu_{l} > 1/n) \ , \nonumber
\end{eqnarray*}
and we prove the following statement:
\begin{theorem}\label{explicit-expression-m1-wep}
The sequence of measures $\tilde{m}_{l}^{(1)}$ converges weakly to a measure $\tilde{m}^{(1)}$, which is
absolutely continuous with respect to $\nu^{0}$ with density $F(\varepsilon)$ given by:
\begin{eqnarray*}
F(\varepsilon) \, = \, \int_{\Lambda_{1}} \ud x  \ \big(e^{\beta(\varepsilon + v(x) - \mu_{\infty})} - 1\big)^{-1} \ .
\end{eqnarray*}
\end{theorem}
\textbf{Proof}: {We follow the line of reasoning of the proof} of Theorem
\ref{Laplace-transform-occupation-measure}.
Let $g_{l}(t; \beta, \mu_{l})$ be the Laplace transform of the measure $\tilde{m}_{l}^{(1)}$:
\begin{eqnarray}\label{Laplace-transform-weak-external-potential}
g_{l}(t; \beta, \mu_{l}) &=& \int_{\mathbb{R}} \,  m_{l}^{(1)} (\ud \varepsilon) \ e^{-t \varepsilon}\\
&=& \sum_{n \geqslant 1} \,  \frac{1}{V_{l}} \, \textrm{Tr} \, e^{-t \, h_{l}^{0}} \, (e^{-n\beta(h_{l}
- \mu_{l})}) \, \mathbf{1} (\mu_{l} \leqslant 1/n) \nonumber
\end{eqnarray}
Again, our aim is to show the uniform convergence of the sum over $n$ with respect to $l$. Let
\begin{eqnarray}\label{definition-coeff-aln-wep}
a_{l}(n) &:=& \frac{1}{V_{l}} \textrm{Tr} \, e^{-t \, h_{l}^{0}} \, e^{ -n \beta(h_{l} - \mu_{l})}
\mathbf{1}(\mu_{l} \leqslant 1/n)\\
&\leqslant& \frac{1}{V_{l}} \textrm{Tr} \, e^{ -n \beta(h_{l} - \mu_{l})} \mathbf{1}(\mu_{l}
\leqslant 1/n) \ . \nonumber
\end{eqnarray}
Then for $\overline{\rho} < \rho_{c}$ we can apply a similar argument as for the random case,
since the estimate $\mu_{l} < \mu_{\infty} / 2 < 0$ still holds, to obtain:
\begin{eqnarray*}
a_{l}(n) \, \leqslant \, e^{n \beta \mu_{\infty} / 2} \int_{[0,\infty)} e^{-\beta \varepsilon} \nu_{l}(\ud \varepsilon) \,
\leqslant \, K_{1} \, e^{n \beta \mu_{\infty} / 2} \ .
\end{eqnarray*}
If $\overline{\rho} \geqslant \rho_{c}$, then $\mu_{l} \leqslant 1/n$ in (\ref{definition-coeff-aln-wep})
implies that:
\begin{eqnarray*}
a_{l}(n) \leqslant e^{\beta}\sum_{i} e^{-n\beta \,E_{i}^{l}} \leqslant \frac{e^{\beta}}{(2 \pi n \beta)^{d/2}}
\int_{\Lambda_{1}} \ud x e^{-n \beta v(x)},
\end{eqnarray*}
where the last estimate can be found in \cite{P} or \cite{VdBL}. Now the uniform convergence
for the sequence
$a_{l}(n)$ follows from (\ref{critical-density-weak-external-potential}), since we assumed that $\rho_{c} < \infty$.
{The latter implies also that for $\overline{\rho}\geq \rho_{c}$, $\mu_{\infty}(\beta,\overline{\rho}) = 0$.}
Thus, we can take the limit of the Laplace transform (\ref{Laplace-transform-weak-external-potential})
term by term, that is:
\begin{eqnarray}\label{Pastur-trick-original-version-weak-external-potential}
\lim_{l \rightarrow \infty} a_{l} (n) &=&\lim_{l \rightarrow \infty} \, \frac{1}{V_{l}} \textrm{Tr} \,
e^{-t \, h_{l}^{0}} \, e^{ -n \beta(h_{l} - \mu_{l})}  \mathbf{1}(\mu_{l} \leqslant 1/n)\\
&=& \lim_{l \rightarrow \infty} \, \frac{1}{V_{l}} \, \int_{\Lambda_{l}} \, \int_{\Lambda_{l}} \ud x \,
\ud x' \, e^{-t \, h_{l}^{0}} (x, x') \, e^{ -n \beta(h_{l} - \mu_{l})} (x',x)\nonumber\\
&=& e^{n\beta\mu_{\infty}} \lim_{l \rightarrow \infty} \, \frac{1}{V_{l}} \, \int_{\Lambda_{l}} \,
\int_{\Lambda_{l}} \ud x \, \ud x' \ \  \frac{e^{-\|x-x'\|^2(1/2n\beta+1/2t)}}{(4\pi^2tn\beta)^{d/2}}
\times \nonumber\\
&\,& \times \int_{\Omega_{(x,x')}^{t}} \, w^{t}(\ud \xi') \,
\chi_{\Lambda_{l},t} (\xi') \int_{\Omega_{(x',x)}^{n\beta}} \, w^{n\beta}(\ud \xi)
\, e^{-\int_{0}^{n\beta} \,
\ud s \, v (\xi(s)/l)} \, \chi_{\Lambda_{l},n\beta} (\xi) \ . \nonumber
\end{eqnarray}
Here we have used the Feynman-Kac representation for free $e^{-t \, h_{l}^{0}} (x,y)$  and  for non-free
$e^{-\beta h_{l}}(x,y)$ Gibbs semi-group kernels, where $w^{T}$ stands for the
\textit{normalized} Wiener measure on the path-space $\Omega_{(x,y)}^{T}$, see Section
\ref{Sect-kinetic-occupation-measure-main-theorem}.

Note that by Lemma \ref{estimate-finite-volume-errors}, which demands only the \textit{non-negativity} of the
potential $v$, we obtain for (\ref{Pastur-trick-original-version-weak-external-potential}) the representation:
\begin{eqnarray}\label{Pastur-trick-Rd-no-paths-restriction-weak-external-potential}
&\,&\lim_{l \rightarrow \infty} \, \frac{1}{V_{l}} \textrm{Tr} \, e^{-t \, h_{l}^{0}} \, e^{ -n \beta(h_{l}
- \mu_{l})}\\
&=& e^{n\beta\mu_{\infty}} \lim_{l \rightarrow \infty} \, \int_{\mathbb{R}^{d}} \ud x\, \frac{1}{V_{l}}
\int_{\Lambda_{l}}  \ud x' \, \frac{e^{-\|x-x'\|^2(1/2n\beta+1/2t)}}{(4\pi^2tn\beta)^{d/2}}
\int_{\Omega_{(x',x)}^{n\beta}} \, w^{n\beta}(\ud \xi) \, e^{-\int_{0}^{n\beta} \,
\ud s \, v (\xi(s)/l)} \ .  \nonumber
\end{eqnarray}

Now we express the trajectories $\xi$ in terms of \textit{Brownian bridges} $\alpha(\tau) \in \tilde{\Omega},
0 \leqslant \tau \leqslant 1$, we denote  the corresponding measure by $D$.
Letting $\tilde{x} = x'/l$, we obtain:
\begin{eqnarray*}
&\,&\lim_{l \rightarrow \infty} \, \frac{1}{V_{l}} \textrm{Tr} \, e^{-t \, h_{l}^{0}} \,
e^{ -n \beta(h_{l} - \mu_{l})}\\
&=& e^{n\beta\mu_{\infty}} \lim_{l \rightarrow \infty} \, \int_{\mathbb{R}^{d}} \ud x\, \int_{\Lambda_{1}}
\ud \tilde{x} \, \frac{e^{-\|x-l\tilde{x}\|^2(1/2n\beta+1/2t)}}{(4\pi^2tn\beta)^{d/2}}  \times\\
&\times&\int_{\tilde{\Omega}} \, D(\ud \alpha) \, \exp \Big(-\int_{0}^{n\beta} \, \ud s \, v [(1-
\frac{s}{n \beta})\tilde{x} + \frac{s}{n \beta} (x/l) + \frac{\sqrt{n \beta}}{l}
\alpha ({s}/{n \beta})]\Big) \ .
\end{eqnarray*}
Since the integration with respect to $x$ is now over the whole space, we let $y = x - l \tilde{x}$ to get
\begin{eqnarray*}
&\,&\lim_{l \rightarrow \infty} \, \frac{1}{V_{l}} \textrm{Tr} \, e^{-t \, h_{l}^{0}} \, e^{ -n \beta(h_{l} -
\mu_{l})}\\
&=& e^{n\beta\mu_{\infty}} \lim_{l \rightarrow \infty} \, \int_{\mathbb{R}^{d}} \ud y\, \int_{\Lambda_{1}}
\ud \tilde{x} \, \frac{e^{-\|y\|^2(1/2n\beta+1/2t)}}{(4\pi^2tn\beta)^{d/2}}  \times\\
&\times&\int_{\tilde{\Omega}} \, D(\ud \alpha) \, \exp \Big(-\int_{0}^{n\beta} \, \ud s\, v \big(\tilde{x} +
\frac{s}{n \beta} (y/l) + \frac{\sqrt{n \beta}}{l} \alpha ({s}/{n \beta}) \big)\Big)\\
&=& e^{n\beta\mu_{\infty}} \, \int_{\mathbb{R}^{d}} \ud y \, \frac{e^{-\|y\|^2(1/2n\beta+1/2t)}}
{(4\pi^2tn\beta)^{d/2}} \, \int_{\Lambda_{1}} \ud \tilde{x} e^{-n \beta v(\tilde{x})} \ \ ,
\end{eqnarray*}
where the last step follows from dominated convergence. Therefore, we obtain by
(\ref{Laplace-transform-weak-external-potential}) the following expression for the limiting Laplace
transform:
\begin{eqnarray*}
\lim_{l \rightarrow \infty} \, g_{l}(t; \beta, \mu_{l}) \, = \, \sum_{n \geqslant 1}
e^{-n\beta(E - \mu_{\infty})}
\frac{1}{(2 \pi (n \beta + t))^{d/2}} \int_{\Lambda_{1}} \ud x e^{-n \beta v(x)},\\
\end{eqnarray*}
It is now straightforward to invert this Laplace transform (for each term of the sum), to find that:
\begin{eqnarray*}
F(E) \ \nu^{0}(\ud E) = \lim_{l \rightarrow \infty} \, \tilde{m}_{l}^{1} (\ud E) \, =
\, \sum_{n \geqslant 1} e^{-n\beta(E - \mu_{\infty})}
\Big( \int_{\Lambda_{1}} \ud x e^{-n \beta v(x)} \Big) \nu^{0}(\ud E)  \ .
\end{eqnarray*}
The Theorem then follows by Fubini's theorem.  \hfill $\square$

\noindent \textbf{Proof of Theorem \ref{existence-occupation-measure-and-BEC-weak-potential}:}
The proof of Theorem \ref{existence-occupation-measure-and-BEC} can be applied directly.
Note that Lemmas
\ref{lower-bound-kineticBEC}, \ref{absolute-continuity-kinetic-occupation-measure} are still valid,
since (as we emphasized in Remark \ref{remark-validity-lemmas}) their proofs require only the
non-negativity
of the external potential. Similarly, Corollary \ref{weight-measure-m1} now can be used directly,
since we
have proved Theorem \ref{explicit-expression-m1-wep}.
\hfill $\square$
\section*{Appendices}
\appendix
\section{Brownian paths}\label{appendix-brownian-path}
\renewcommand{\theequation}{\Alph{section}.\arabic{equation}}
\renewcommand{\thelemma}{\Alph{section}.\arabic{lemma}}
\setcounter{section}{1}
\setcounter{equation}{0}
\setcounter{theorem}{0}
\setcounter{lemma}{0}
In this section, we first give an upper estimate of the probability of a Brownian path to leave some spatial
domain, cf. e.g. \cite{MMP} and the references quoted therein.
\begin{lemma}\label{upper-bound-brownian-bridge}
Let the set
$$\Omega_{(x,x')}^{T} := \{ \xi(\tau) : \xi(0) = x, \, \xi(T) = x' \}$$
be continuous trajectories from $x$ to $x'$ with the proper time $0\leq\tau\leq T$, and with
the normalized Wiener measure $w^{T}$ on it. Let $x,x'$ be in $\Lambda_{l}$, and
$\chi_{\Lambda_{l}, T} (\xi)$ the
characteristic function over $\Omega_{(x,x')}^{T}$
of trajectories $\xi$ staying in $\Lambda_{l}$ for all $0 \leqslant \tau \leqslant T$. Then one gets the
estimate:
\begin{eqnarray}\label{exit}
\int_{\Omega_{(x,x')}^{T}} \, w^{T}(\ud \xi) \Big(1-\chi_{\Lambda_{l}, T} (\xi) \Big) \,
\leqslant \, e^{- C(T) \big(\min \{d(x,\partial\Lambda_{l}),  d(x',\partial\Lambda_{l})\}\big)^{2}} \ .
\end{eqnarray}
\end{lemma}
\textbf{Proof:}
Define a \textit{Brownian bridge} $\alpha(s), 0\leqslant s \leqslant 1$ by:
\begin{eqnarray*}
\xi(t) = (1-{\tau}/{T})\, x + {\tau}/{T} \ x' + \sqrt{T}\, \alpha({\tau}/{T}).
\end{eqnarray*}
Let us consider first the one dimensional case, i.e. $\Lambda_{l} = \left[-l/2, l/2\right]$. Without
loss of generality, we can assume that:
\begin{eqnarray*}
d(x,\partial\Lambda_{l}) \, \leqslant \, d(x',\partial\Lambda_{l}).
\end{eqnarray*}

Suppose that  $x > 0$, then we have:
\begin{eqnarray*}
-x \leqslant x' \leqslant x \quad \textrm{and} \quad d(x,\partial\Lambda_{l}) = l/2-x
\end{eqnarray*}
Assume that the path $\xi$ leaves the box on the right-hand side. Then, for some $t$, we have:
\begin{eqnarray}\label{brownian-bridge-1D-1}
\xi(t)&>& \frac{l}{2}\nonumber\\
\alpha({t}/{T}) &>& \frac{1}{\sqrt{T}} \Big( \frac{l}{2} + ({t}/{T}-1) x - \frac{t}{T} x' \Big)\nonumber\\
\alpha({t}/{T}) &>& \frac{1}{\sqrt{T}} \Big( \frac{l}{2} + ({t}/{T}-1) x - \frac{t}{T} x \Big)\, =
\, \frac{1}{\sqrt{T}} d(x,\partial\Lambda_{l})
\end{eqnarray}
The case, when $\xi$ leaves the box on the left-hand side can be treated similarly.

Let $x < 0$, then we have:
\begin{eqnarray*}
x \leqslant x' \leqslant -x \quad \textrm{and} \quad d(x,\partial\Lambda_{l}) = l/2+x
\end{eqnarray*}
Again, assume that the path leaves the box on the right hand-side. Then, for some $t$, we have:
\begin{eqnarray}\label{brownian-bridge-1D-2}
\xi(t)&>& \frac{l}{2}\nonumber\\
\alpha({t}/{T}) &>& \frac{1}{\sqrt{T}} \Big( \frac{l}{2} + ({t}/{T}-1) x - \frac{t}{T} x' \Big)\nonumber\\
\alpha({t}/{T}) &>& \frac{1}{\sqrt{T}} \Big( \frac{l}{2} - ({t}/{T}-1) x' - \frac{t}{T} x' \Big)\,
\geqslant \, \frac{1}{\sqrt{T}} d(x,\partial\Lambda_{l})
\end{eqnarray}
The case, when $\xi$ leaves the box on the left hand-side can be considered similarly. The relations
(\ref{brownian-bridge-1D-1}), (\ref{brownian-bridge-1D-2}) imply that if
$\xi$ leaves the box $\Lambda_{l}$ in one dimension, then the Brownian bridge $\alpha$ must satisfy the
inequality:
\begin{eqnarray}\label{sup-brownian-bridge-1D}
\sup_{t} \, |\alpha({t}/{T})| \,\, > \,\, C(T) \min \{d(x,\partial\Lambda_{l}),  d(x',\partial\Lambda_{l})\},
\end{eqnarray}
for some constant $C(T)$.

This observation can easily be extended to higher dimensions, when $x:= (x_{1}, \dots, x_{d})$
and $\alpha(s) := (\alpha_{1}(s), \dots, \alpha_{d}(s))$. Now, if $\xi$ leaves the ($d$-dimensional)
box $\Lambda_{l}$,
there exists at least one $i$ such that similar to (\ref{sup-brownian-bridge-1D}):
\begin{eqnarray*}
\sup_{t} \, |\alpha_{i}({t}/{T})| \,\, > \,\, C(T) \min \{d(x_{i},\partial_{i}\Lambda_{l}),
d(x_{i}',\partial_{i}\Lambda_{l})\},
\end{eqnarray*}
where we denote $d(x_{i},\partial_{i}\Lambda_{l}) := \min \{l/2-x_{i}, l/2+x_{i}\}$. Now, since $\Lambda_{l}$
are cubes,
we get $d(x_{i},\partial_{i}\Lambda_{l} \geqslant d(x,\partial\Lambda_{l}$ for any $x \in \Lambda_{l}$.
Then we obtain:
\begin{eqnarray}\label{sup-brownian-bridge}
\|\alpha({t}/{T})\| \, &>& \,  |\alpha_{i}({t}/{T})|, \qquad i = 1,\dots,d  \ , \nonumber\\
\sup_{t} \|\alpha({t}/{T})\|  \,&>& \, \max_{i}\sup_{t}  |\alpha_{i}({t}/{T})|  \ , \nonumber\\
\sup_{t} \|\alpha({t}/{T})\|  \,&>& \, C(T) \min \{d(x_{i},\partial_{i}\Lambda_{l}),
d(x_{i}',\partial_{i}\Lambda_{l})\}
\nonumber \\
\, &\geqslant& \, C(T) \min \{d(x,\partial\Lambda_{l}),  d(x',\partial\Lambda_{l})\} \ .
\end{eqnarray}
Therefore, the probability for the path $\xi$ to leave the box is dominated by the probability for the
one-dimensional Brownian bridge $\alpha$  to satisfy (\ref{sup-brownian-bridge}). The latter we can
estimate using
the following result from \cite{MMP}:
\begin{eqnarray*}
\mathbb{P} \Big( \sup_{s} \alpha(s) \, > \, x\Big) \geqslant A e^{-Cx^{2}}
\end{eqnarray*}
valid for some positive constants $A,C$, which implies the bound (\ref{exit}).
\hfill $\square$

Now we establish a result, that we use in the proof of Theorem \ref{Laplace-transform-occupation-measure}:
\begin{lemma}\label{estimate-finite-volume-errors}
Let $K^{t}_{\omega,l}(x,x')$, $K^{t}_{0,l}(x,x')$, $K^{t}_{0}(x,x')$ be the kernels of operators
$\exp (-t h_{l}^{\omega})$, $\exp (-t h_{l}^{0})$, and $\exp (- t \Delta/2)$ respectively. Then
\begin{eqnarray}\label{limiting identity}
&&\lim_{l \rightarrow \infty} \, \frac{1}{V_{l}} \, \int_{\Lambda_{l}} \, \int_{\Lambda_{l}} \ud x \,
\ud x' \,
K^{t}_{0,l}(x,x') \, K^{n\beta}_{\omega,l}(x',x)  \\
&&\hskip 3cm =\lim_{l \rightarrow \infty} \, \int_{\mathbb{R}^{d}} \ud x\,
\frac{1}{V_{l}}\int_{\Lambda_{l}}  \ud x' \, K^{t+n\beta}_{0}(x,x')  \int_{\Omega_{(x',x)}^{n\beta}} \,
w^{n\beta}(\ud \xi) \, e^{-\int_{0}^{n\beta} \, \ud s  \, v^{\omega} (\xi(s))} \,  . \nonumber
\end{eqnarray}
\end{lemma}
\textbf{Proof}: By the Feynman-Kac representation, we obtain:
\begin{eqnarray*}
&&\lim_{l \rightarrow \infty} \, \frac{1}{V_{l}} \, \int_{\Lambda_{l}} \, \int_{\Lambda_{l}} \ud x \, \ud x'
\, K^{t}_{0,l}(x,x') \, K^{n\beta}_{\omega,l}(x',x) =\\
&&\lim_{l \rightarrow \infty} \, \frac{1}{V_{l}} \, \int_{\Lambda_{l}} \, \int_{\Lambda_{l}} \ud x \, \ud x'
\, \frac{e^{-\|x-x'\|^2(1/2n\beta+1/2t)}}{(4\pi^2tn\beta)^{d/2}} \int_{\Omega_{(x',x)}^{n\beta}} \,
w^{n\beta}(\ud \xi) \, e^{-\int_{0}^{n\beta} \, \ud s \, v^{\omega} (\xi(s))} \,
\chi_{\Lambda_{l},n\beta}
(\xi) \times \\
&&\times \int_{\Omega_{(x,x')}^{t}} \, w^{t}(\ud \xi') \, \chi_{\Lambda_{l},t} (\xi') \ .
\end{eqnarray*}
To eliminate the characteristic functions restricting the paths $\xi, \xi'$ in the last integral, we shall
use Lemma \ref{upper-bound-brownian-bridge}.
First, we estimate the error $\gamma(d)$ when we remove the restriction on the path $\xi$:
\begin{eqnarray}\label{Pastur-trick-Rd-error-paths}
\gamma(d) &:=&   \lim_{l \rightarrow \infty} \, \frac{1}{V_{l}} \, \int_{\Lambda_{l}}  \ud x\,
\int_{\Lambda_{l}}
\ud x' \, \frac{e^{-\|x-x'\|^2(1/2n\beta+1/2t)}}{(4\pi^2tn\beta)^{d/2}} \times \\
&\,& \times \int_{\Omega_{(x',x)}^{n\beta}} \, w^{n\beta}(\ud \xi) \, e^{-\int_{0}^{n\beta} \,
\ud s \, v^{\omega}
(\xi(s))} \, \big(1-\chi_{\Lambda_{l},n\beta} (\xi)\big)\int_{\Omega_{(x,x')}^{t}} \, w^{t}(\ud \xi')
\, \chi_{\Lambda_{l},t} (\xi')\nonumber\\
&\leqslant&   \lim_{l \rightarrow \infty} \, \frac{1}{V_{l}} \, \int_{\Lambda_{l}} \ud x\, \int_{\Lambda_{l}}
\ud x' \, \frac{e^{-\|x-x'\|^2(1/2n\beta+1/2t)}}{(4\pi^2tn\beta)^{d/2}} \int_{\Omega_{(x',x)}^{n\beta}} \,
w^{n\beta}(\ud \xi) \, \big(1-\chi_{\Lambda_{l},n\beta} (\xi)\big)\nonumber\\
&\leqslant& \lim_{l \rightarrow \infty} \, \frac{1}{V_{l}} \, \int_{\Lambda_{l}} \ud x\, \int_{\Lambda_{l}}
\ud x' \, \mathbb{I}\{d(x,\partial\Lambda_{l}) \, > \, d(x',\partial\Lambda_{l})\} \,
\frac{e^{-\|x-x'\|^2(1/2n\beta+1/2t)}}{(4\pi^2tn\beta)^{d/2}} \times \nonumber\\
&\times &\int_{\Omega_{(x',x)}^{n\beta}} \, w^{n\beta}(\ud \xi) \, \big(1-\chi_{\Lambda_{l},n\beta}
(\xi)\big)\nonumber\\
&+& \lim_{l \rightarrow \infty} \, \frac{1}{V_{l}} \, \int_{\Lambda_{l}} \ud x\, \int_{\Lambda_{l}}
\ud x' \, \mathbb{I}\{d(x,\partial\Lambda_{l} \, \leqslant \, d(x',\partial\Lambda_{l})\} \ ,
\frac{e^{-\|x-x'\|^2(1/2n\beta+1/2t)}}{(4\pi^2tn\beta)^{d/2}} \times \nonumber\\
&\times &\int_{\Omega_{(x',x)}^{n\beta}} \, w^{n\beta}(\ud \xi) \, \big(1-\chi_{\Lambda_{l},n\beta}
(\xi)\big)\nonumber\\
&\leqslant&   \lim_{l \rightarrow \infty} \, \frac{1}{V_{l}} \, \int_{\Lambda_{l}} \ud x\, \int_{\Lambda_{l}}
\ud x' \,K_{0}^{t}(x,x')K_{0}^{n\beta}(x',x) e^{- C(n\beta) (d(x',\partial\Lambda_{l})^{2}}\nonumber\\
&+& \lim_{l \rightarrow \infty} \, \frac{1}{V_{l}} \, \int_{\Lambda_{l}} \ud x\, \int_{\Lambda_{l}}
\ud x' K_{0}^{t}(x,x')K_{0}^{n\beta}(x',x) e^{- C(n\beta) (d(x,\partial\Lambda_{l})^{2}}\nonumber
\end{eqnarray}
where the last step is due to Lemma \ref{upper-bound-brownian-bridge}. Since all integrands are
positive, we can
extend one of the spatial integrations to the whole space,  and hence we get:
\begin{eqnarray*}
\gamma(d) &\leqslant&   \lim_{l \rightarrow \infty} \, \frac{1}{V_{l}} \, \int_{\mathbb{R}^{d}} \ud x\,
\int_{\Lambda_{l}} \ud x' \,K_{0}^{t}(x,x')K_{0}^{n\beta}(x',x) e^{- C(n\beta)
(d(x',\partial\Lambda_{l})^{2}}\\
&+& \lim_{l \rightarrow \infty} \, \frac{1}{V_{l}} \, \int_{\Lambda_{l}} \ud x\, \int_{\mathbb{R}^{d}}
\ud x' K_{0}^{t}(x,x')K_{0}^{n\beta}(x',x) e^{- C(n\beta) (d(x'\partial\Lambda_{l})^{2}}\\
&\leqslant&  \lim_{l \rightarrow \infty} \, \frac{1}{V_{l}} \, K_{0}^{t+n\beta} \int_{\Lambda_{l}}
\ud x'  e^{- C(n\beta) (d(x',\partial\Lambda_{l})^{2}}\\
&+& \lim_{l \rightarrow \infty} \, \frac{1}{V_{l}} \, K_{0}^{t+n\beta} \int_{\Lambda_{l}} \ud x\,
e^{- C(n\beta) (d(x'\partial\Lambda_{l})^{2}}\, ,
\end{eqnarray*}
{where we have used the notation $K_{0}^{t+n\beta} := K_{0}^{t+n\beta}(x,x)$ since these are independent of $x$.}
Finally, using the fact that the boxes $\Lambda_{l}$ are cubes of side $l$, we obtain:
\begin{eqnarray*}
\gamma(d) \leqslant  \lim_{l \rightarrow \infty} \frac{K_{0}^{t+n\beta}}{l} \int_{-l/2}^{l/2} \ud x'
e^{- C(n\beta) (l/2-x')^{2}} +  \lim_{l \rightarrow \infty} \frac{K_{0}^{t+n\beta}}{l} \int_{-l/2}^{l/2}
\ud x e^{- C(n\beta) (l/2-x)^{2}}\, = \, 0
\end{eqnarray*}
We can estimate the error estimate due to the removal of the characteristic function for $\xi'$ in
(\ref{Pastur-trick-original-version}) in the same way. Therefore, we get:
\begin{eqnarray}\label{Pastur-trick-without-path-restriction}
&\,& \lim_{l \rightarrow \infty} \, \frac{1}{V_{l}} \, \int_{\Lambda_{l}} \, \int_{\Lambda_{l}}
\ud x \, \ud x'
\, \frac{e^{-\|x-x'\|^2(1/2n\beta+1/2t)}}{(4\pi^2tn\beta)^{d/2}}\times \\
&\,&\times \int_{\Omega_{(x',x)}^{n\beta}} \, w^{n\beta}(\ud \xi) \, e^{-\int_{0}^{n\beta} \,
\ud s \, v^{\omega}
(\xi(s))} \, \chi_{\Lambda_{l},n\beta} (\xi)\int_{\Omega_{(x,x')}^{t}} \, w^t(\ud \xi') \,
\chi_{\Lambda_{l},t} (\xi') \nonumber\\
&=& \lim_{l \rightarrow \infty} \frac{1}{V_{l}} \int_{\Lambda_{l}} \int_{\Lambda_{l}} \ud x \ud x'
\frac{e^{-\|x-x'\|^2(1/2n\beta+1/2t)}}{(4\pi^2tn\beta)^{d/2}}\int_{\Omega_{(x',x)}^{n\beta}}
w^{n\beta}(\ud \xi) e^{-\int_{0}^{n\beta} \, \ud s \, v^{\omega} (\xi(s))} \int_{\Omega_{(x,x')}^{p}}
w^{n\beta}(\ud \xi') \ .\nonumber
\end{eqnarray}
Now we show that one can replace the first integration over the box $\Lambda_{l}$ by one over the
whole space. Let $\tilde{\gamma}(d)$ be the error caused by this substitution. Then by the positivity of the
random potential we get the estimate:
\begin{eqnarray}\label{error-Pastur-trick-Lambda->Rd}
\tilde{\gamma}(d) &:=& \lim_{l \rightarrow \infty} \, \frac{1}{V_{l}} \,
\int_{{\mathbb{R}^d} \setminus \Lambda_{l}} \ud x\,
\int_{\Lambda_{l}}  \ud x' \, \frac{e^{-\|x-x'\|^2(1/2n\beta+1/2t)}}{(4\pi^2tn\beta)^{d/2}} \times \\
&\,&\times\int_{\Omega_{(x',x)}^{n\beta}} \, w^{n\beta}(\ud \xi) \, e^{-\int_{0}^{n\beta} \,
\ud s \, v^{\omega} (\xi(s)+x')} \int_{\Omega_{(x,x')}^{t}} \, w^{n\beta}(\ud \xi') \,\nonumber\\
&\leqslant&  \lim_{l \rightarrow \infty} \, \frac{1}{V_{l}} \,
\int_{{\mathbb{R}^d} \setminus \Lambda_{l}} \ud x \,
\int_{\Lambda_{l}}  \ud x' \, \frac{e^{-\|x-x'\|^2(1/2n\beta+1/2t)}}{(4\pi^2tn\beta)^{d/2}} \ \ .\nonumber
\end{eqnarray}
In the one-dimensional case the estimate of the error term (\ref{error-Pastur-trick-Lambda->Rd}) takes the
form:
\begin{eqnarray}\label{error-Pastur-trick-Lambda->Rd-1d}
\tilde{\gamma}(1) & \leqslant &  \lim_{l \rightarrow \infty} \, \frac{1}{l} \, \int_{-\infty}^{-l/2} \,\ud x
\int_{-l/2-x}^{l/2-x}  \textrm{d}y \, \frac{e^{-y^2(1/2n\beta+1/2t)}}{(4\pi^2tn\beta)^{1/2}}\\
&+&  \lim_{l \rightarrow \infty} \, \frac{1}{l} \, \int_{l/2}^{\infty} \,\ud x \int_{-l/2-x}^{l/2-x}
\textrm{d}y \, \frac{e^{-y^2(1/2n\beta+1/2t)}}{(4\pi^2tn\beta)^{1/2}}.\nonumber
\end{eqnarray}
For the first term one gets:
\begin{eqnarray*}
&& \lim_{l \rightarrow \infty} \, \frac{1}{l} \, \int_{-\infty}^{-l/2} \,\ud x \int_{-l/2-x}^{l/2-x}
\textrm{d}y \, \frac{e^{-y^2(1/2n\beta+1/2t)}}{(4\pi^2tn\beta)^{1/2}} \ = \\
&&  \lim_{l \rightarrow \infty} \, \frac{1}{l} \int_{0}^{l} \,\textrm{d}y \frac{e^{-y^2(1/2n\beta+1/2t)}}
{(4\pi^2tn\beta)^{1/2}}\int_{-l/2-y}^{l/2}  \ud x
+  \lim_{l \rightarrow \infty} \, \frac{1}{l} \int_{l}^{\infty} \,\textrm{d}y
\frac{e^{-y^2(1/2n\beta+1/2t)}}
{(4\pi^2tn\beta)^{1/2}}\int_{-l/2-y}^{l/2-y}  \ud x
= 0 \ .
\end{eqnarray*}
One obtains a similar identity for the second-term in (\ref{error-Pastur-trick-Lambda->Rd-1d}).
Direct calculation shows that, the error term for higher dimensions (\ref{error-Pastur-trick-Lambda->Rd})
reduces to a product of one-dimensional terms (\ref{error-Pastur-trick-Lambda->Rd-1d}). Then
(\ref{Pastur-trick-without-path-restriction}) gives:
\begin{eqnarray}\label{Pastur-trick-Rd-no-paths-restriction}
&\,& \lim_{l \rightarrow \infty} \, \frac{1}{V_{l}} \, \int_{\Lambda_{l}} \, \int_{\Lambda_{l}} \ud x \,
\ud x' \, \frac{e^{-\|x-x'\|^2(1/2n\beta+1/2t)}}{(4\pi^2tn\beta)^{d/2}} \times \\
&\,& \times \int_{\Omega_{(x',x)}^{n\beta}} \, w^{n\beta}(\ud \xi) \, e^{-\int_{0}^{n\beta} \,
\ud s \, v^{\omega} (\xi(s))} \, \chi_{\Lambda_{l},n\beta} (\xi)\int_{\Omega_{(x,x')}^{t}} \,
w^{t}(\ud \xi') \, \chi_{\Lambda_{l},t} (\xi') \nonumber\\
&=& \lim_{l \rightarrow \infty} \, \int_{\mathbb{R}^{d}} \ud x\, \frac{1}{V_{l}}\int_{\Lambda_{l}}
\ud x' \, \frac{e^{-\|x-x'\|^2(1/2n\beta+1/2t)}}{(4\pi^2tn\beta)^{d/2}}\int_{\Omega_{(x',x)}^{n\beta}} \,
w^{n\beta}(\ud \xi) \, e^{-\int_{0}^{n\beta} \, \ud s \, v^{\omega} (\xi(s))} \ , \nonumber
\end{eqnarray}
which finishes the proof of (\ref{limiting identity}).
\hfill $\square$
\section{Some probabilistic estimates}\label{appendix-probab-estimates}
\renewcommand{\theequation}{\Alph{section}.\arabic{equation}}
\renewcommand{\thelemma}{\Alph{section}.\arabic{lemma}}
\setcounter{section}{2}
\setcounter{equation}{0}
\setcounter{theorem}{0}
\setcounter{lemma}{0}
First we recall the assumptions on the random potential $v^{\omega}$ used in \cite{KM}, and which we also
adopt in this paper:
\begin{enumerate}
  \item
  \begin{enumerate}
    \item On the probability space $(\Omega, \mathcal{F}, \mathbb{P})$ there exist a group of measure-preserving
metrically transitive transformations $\{T_{p}\}_{p \in \mathbb{R}^{d}}$ of $\Omega$, such that
$v^{\omega}(x+p) = v^{T_{p}\omega}(x)$ for all $x, p \in \mathbb{R}^{d}$;
    \item $\mathbb{E}_{\omega} \{ \int_{\Lambda_1} \ud x \, |v^{\omega}(x)|^{\kappa}\} < \infty$, where
$\kappa > \max(2,d/2)$.
  \end{enumerate}
  \item
  For any $\Lambda \subset \mathbb{R}^{d}$, let $\Sigma_{\Lambda}$ be the $\sigma$-algebra generated by
the random field $v^{\omega}(x), x \in \Lambda$. For any two arbitrary random variables on
$\Omega$, $f$,$g$ satisfying (i)\ $|g|_{\infty} < \infty$, $\mathbb{E}_{\omega}\{|f|\} < \infty$ and (ii)\ the
function $g$ is $\Sigma_{\Lambda_{1}}$-measurable, the function $f$ is
$\Sigma_{\Lambda_{2}}$-measurable,
where $\Lambda_{1}, \Lambda_{2}$ are disjoint bounded subsets of $\mathbb{R}^{d}$,
  the following holds
\begin{eqnarray*}
|\mathbb{E}\{|f.g|\} \, - \, \mathbb{E}\{|f|\} \mathbb{E}\{|g|\}| \, \leqslant \, |g|_{\infty} \,
\mathbb{E}\{|f|\} \, \phi(d(\Lambda_{1}, \Lambda_{2}))
\end{eqnarray*}
with $\phi(x) \rightarrow 0$ as $x \rightarrow \infty$, and $d(\Lambda_{1}, \Lambda_{2})$ the
Euclidean distance between $\Lambda_{1}$ and $\Lambda_{2}$.
\end{enumerate}

After recalling these conditions, we can give a sketch of the proof of Lemma \ref{upper-bound-eigenvalue1}.

Let $h^{\omega, N}_{l}$ to be the Schr\"{o}dinger operator (\ref{Schrodinger-operator}), with
Neumann boundary conditions instead of Dirichlet, and denote by $\{E_{i}^{\omega,l,N},
\phi_{i}^{\omega,l,N} \}_{i \geqslant 1}$ its ordered eigenvalues (including degeneracy) and the
corresponding eigenvectors. Similarly we define the kinetic energy operator $h^{0, N}_{l}$ with the same
boundary condition, and denote by $\{\varepsilon_{k}^{l,N}, \psi_{k}^{l,N}\}_{k\geqslant 1}$ its
ordered eigenvalues (including degeneracy) and corresponding eigenvectors. The following result is
due to Thirring, see \cite{T}:
\begin{lemma}\label{Thirring-estimate-first-eigenvalue}
Let $v_{\lambda,\alpha}^{\omega} := v^{\omega} + \lambda \, \alpha$, for $\lambda, \alpha > 0$. Then,
\begin{eqnarray*}
E_{1}^{\omega,l,N} \, \geqslant \, - \lambda \alpha \, + \, \min \Big\{ \varepsilon_{2}^{l,N} ,
\left[\frac{1}{V_{l}} \int_{\Lambda_{l}} \ud x \,  (v_{\lambda,\alpha}^{\omega}(x))^{-1} \right]^{-1}
\Big\} \ \ .
\end{eqnarray*}
\end{lemma}
\textbf{Proof}:
Let $P$ to be an orthogonal projection in $\mathscr{H}_{l} $. Then for any vector
$\phi$ from the intersection $Q(v_{\lambda,\alpha}^{\omega})
\bigcap Q((v_{\lambda,\alpha}^{\omega})^{1/2} P (v_{\lambda,\alpha}^{\omega})^{1/2})$, we have:
\begin{eqnarray*}
(\phi, v_{\lambda,\alpha}^{\omega} \phi) &=& ((v_{\lambda,\alpha}^{\omega})^{1/2} \phi,
(v_{\lambda,\alpha}^{\omega})^{1/2} \phi)\\
&=& ((v_{\lambda,\alpha}^{\omega})^{1/2} \phi, P (v_{\lambda,\alpha}^{\omega})^{1/2} \phi) \, +
\, ((v_{\lambda,\alpha}^{\omega})^{1/2} \phi, (1-P) (v_{\lambda,\alpha}^{\omega})^{1/2} \phi)\\
&\geqslant& ((v_{\lambda,\alpha}^{\omega})^{1/2} \phi, P (v_{\lambda,\alpha}^{\omega})^{1/2} \phi),
\end{eqnarray*}
and therefore,
\begin{eqnarray}\label{Thirring-proof-1}
-\shalf \Delta_{N} \, + \, v_{\lambda,\alpha}^{\omega} \, \geqslant \, -\shalf\Delta_{N} \, +
\, (v_{\lambda,\alpha}^{\omega})^{1/2} P (v_{\lambda,\alpha}^{\omega})^{1/2},
\end{eqnarray}
in the quadratic-form sense. Let us choose:
\begin{eqnarray*}
P \, := \, (v_{\lambda,\alpha}^{\omega})^{-1/2} \tilde{P} \, \big( (\psi_{1}^{l,N},
(v_{\lambda,\alpha}^{\omega})^{-1} \psi_{1}^{l,N}) \big)^{-1}\, \tilde{P} (v_{\lambda,\alpha}^{\omega})^{-1/2},
\end{eqnarray*}
where $\tilde{P}$ is the orthogonal projection onto the subspace spanned by the
vector $\psi_{1}^{l,N}$.
It can be easily checked that $P$ is an orthogonal projection. Applying (\ref{Thirring-proof-1}) to the
function $\phi_{1}^{\omega, l, N}$ one gets:
\begin{eqnarray*}
E_{1}^{\omega,l, N} \, + \, \lambda \alpha &\geqslant& (\phi_{1}^{\omega,l,N}, (-\shalf\Delta_{N})
\phi_{1}^{\omega,l,N}) \, + \, |(\phi_{1}^{\omega,l,N}, \psi_{1}^{l,N})|^{2} \,
\big( \psi_{1}^{l,N}, (v_{\lambda,\alpha}^{\omega})^{-1} \psi_{1}^{l,N}\big)^{-1}\\
&\geqslant& \sum_{k \geqslant 1} \, |(\phi_{1}^{\omega,l,N}, \psi_{k}^{l,N})|^{2} \,
\varepsilon_{k}^{l,N} \, + \,  |(\phi_{1}^{\omega,l,N}, \psi_{1}^{l,N})|^{2} \,
\left[ \frac{1}{V_{l}}\int_{\Lambda_{l}} \, \ud x \, (v_{\lambda,\alpha}^{\omega}(x))^{-1} \right]^{-1}.
\end{eqnarray*}
But since the Neumann boundary conditions imply that $\varepsilon_{1}^{l,N} = 0$, we obtain
\begin{eqnarray*}
E_{1}^{\omega,l, N} \, + \, \lambda \alpha \, \geqslant \,  (1 - |(\phi_{1}^{\omega,l,N}, \psi_{1}^{l,N})|^{2})
\, \varepsilon_{2}^{l,N} \, + \, |(\phi_{1}^{\omega,l,N}, \psi_{1}^{l,N})|^{2} \,
\left[ \frac{1}{V_{l}}\int_{\Lambda_{l}} \, \ud x \, (v_{\lambda,\alpha}^{\omega}(x))^{-1} \right]^{-1} \ .
\end{eqnarray*}
To finish the proof, we have to study separately the two cases, namely, $\varepsilon_{2}^{l,N}$ less than and
greater than $\left[ \frac{1}{V_{l}}\int_{\Lambda_{l}} \, \ud x \, (v_{\lambda,\alpha}^{\omega}(x))^{-1} \right]^{-1}$.
\hfill $\square$

\noindent \textbf{Proof of Lemma \ref{upper-bound-eigenvalue1}}:
By Lemma \ref{Thirring-estimate-first-eigenvalue},with $\lambda = B/l^2$ and $\alpha$ as defined in
assumptions, i.e. for $B=\pi/(1+\alpha)$, $\alpha >p/(1-p)$, we have:
\begin{eqnarray*}
 E_{1}^{\omega,l, N} \, \geqslant \, - \frac{\alpha B}{l^{2}} + \min ({\pi}/{l^{2}}, {1}/{X_{l}}) \ ,\\
 \textrm{where} \,\,\, X_{l}^\omega:= \frac{1}{V_{l}} \, \int_{\Lambda_{l}} \ud x \,\frac{1}{v^{\omega}(x) +
 B\alpha/l^{2}} \ .
\end{eqnarray*}
Therefore,
\begin{eqnarray*}
E_{1}^{\omega,l, N} - \frac{B}{l^{2}} \, \geqslant \, - \frac{\pi}{l^{2}} + \min ({\pi}/{l^{2}},
{1}/{X_{l}^\omega}) \ .
\end{eqnarray*}
Hence, the inequality $E_{1}^{\omega,l, N} < {B}/{l^{2}}$ implies that $X_{l}^\omega > l^{2}/\pi$ and
consequently:
\begin{eqnarray}\label{estim1A}
\mathbb{P} ( E_{1}^{\omega,l, N} < {B}/{l^{2}} ) \, \leqslant \, \mathbb{P} ( X_{l}^\omega > l^{2}/\pi ) \ .
\end{eqnarray}
Define a random variable $Y_{l}^{\omega} (\delta):= {V_{l}}^{-1} \, \int_{\Lambda_{l}} \ud x \ \delta/( v ^{\omega}(x) +
\delta)$,
which is an increasing function of $\delta$. Then for the left-hand side of (\ref{estim1A}) one gets the estimate:
\begin{eqnarray*}
\mathbb{P} ( E_{1}^{\omega, l, N} < {B}/{l^{2}} ) \, \leqslant \, \mathbb{P} \Big( Y_{l}^{\omega}({B\alpha}/{l^{2}})>
\frac{\alpha}{1+\alpha} \Big) \ .
\end{eqnarray*}
By Lemma 2 in \cite{KM}, we know that for any positive $\delta$ the random variables $\{Y_{l}^{\omega}(\delta)\}_{l}$,
converges \textit{geometrically} to a limit $Y_{\infty} (\delta)$ as $l \rightarrow \infty$, that is,
for any $\epsilon > 0$, there exists a constant $M(\delta, \epsilon)$
such that
\begin{eqnarray*}
\mathbb{P} ( |Y_{l}^{\omega} (\delta) - Y_{\infty} (\delta)| > \epsilon/2) \, \leqslant \,
e^{- M(\delta, \epsilon) \, V_{l}}
\end{eqnarray*}
for $l$ sufficiently large. By the ergodic theorem $Y_{\infty} (\delta)$ is non-random and can be expressed as:
\begin{eqnarray*}
Y_{\infty} (\delta) \, = \, \mathbb{E}_{\omega} \left(\frac{\delta}{v^{\omega} (0)+\delta}\right) \ ,
\end{eqnarray*}
which is again a monotonic function of $\delta \geq 0$.
Notice that by condition (ii), Section \ref{model-notations}, we have $\lim_{\delta \rightarrow 0} Y_{\infty} (\delta) = p$.

Choose $\epsilon > 0$ such that $p + \epsilon < {\alpha}/{(1+\alpha)}$. Then we have
\begin{eqnarray*}
\mathbb{P} ( E_{1}^{\omega, l, N} < \frac{B}{l^{2}} ) \, \leqslant \, \mathbb{P} \Big( Y_{l}^{\omega}
({B\alpha}/{l^{2}})> p + \epsilon \Big)
\end{eqnarray*}
Now we choose $\delta$ such that
\begin{eqnarray*}
Y_{\infty} (\delta) - p  < \epsilon/2 \ ,
\end{eqnarray*}
and let $l_{0}$ be defined by  $\delta = B \alpha/l_0^2$. Then for any
$l > l_{0}$ we have:
\begin{eqnarray*}
\mathbb{P} ( E_{1}^{\omega, l, N} < {B}/{l^{2}} ) &\leqslant& \mathbb{P} \Big( Y_{l}^{\omega}
({B\alpha}/{l^{2}})> p +
\epsilon \Big) \,\leqslant\mathbb{P} \Big( Y_{l}^{\omega} (\delta) - p >  \epsilon \Big)\\
&\leqslant & \mathbb{P}
\Big( |Y_{l}^{\omega} (\delta) - Y_{\infty} (\delta)| >  \epsilon/2 \Big) \, \leqslant \, e^{-M(\delta,
\epsilon) \, V_{l}} \ .
\end{eqnarray*}
\hfill $\square$
}

\section*{Acknowledgments}

{One of the authors (Th.Jaeck) is supported by funding from the UCD Ad Astra Research Scholarship.}


\end{document}